\documentclass{aa}
\usepackage{graphicx}
\usepackage{txfonts}
\usepackage{hyperref}
\usepackage{dcolumn}

\def\cratio{$^{12}$C/$^{13}$C}
\def\nratio{$^{14}$N/$^{15}$N}

\newcommand{\asec}{$^{\prime\prime}$}
\newcommand{\GG}[1]{}

\newcolumntype{.}{D{.}{.}{-1}}

\begin{document}

   \title{CHEMOUT: CHEMical complexity in star-forming regions of the OUTer Galaxy }

  \subtitle{III. Nitrogen isotopic ratios in the outer Galaxy}

   \author{L. Colzi\inst{1}
         \and
         D. Romano\inst{2}
        \and 
         F. Fontani\inst{3}
         \and
         V. M. Rivilla\inst{1}
         \and
         L. Bizzocchi\inst{4}
          \and
          M. T. Beltrán\inst{3}
          \and
          P. Caselli\inst{5}
          \and
         D. Elia\inst{6}
         \and
         L. Magrini\inst{3}
                         }
          
   \institute{Centro de Astrobiología (CAB), CSIC-INTA, Ctra. de Ajalvir Km. 4, 28850, Torrejón de Ardoz, Madrid, Spain \\
              \email{lcolzi@cab.inta-csic.es}
             \and 
               INAF, Osservatorio di Astrofisica e Scienza dello Spazio, Via Gobetti 93/3, 40129, Bologna, Italy
               \and
            INAF-Osservatorio Astrofisico di Arcetri, Largo E. Fermi 5, I-50125, Florence, Italy
            \and
            Department of Chemistry “Giacomo Ciamician”, University of Bologna, Via F. Selmi 2, Bologna, 40126, Italy 
            \and
            Centre for Astrochemical Studies, Max-Planck-Institute for Extraterrestrial Physics, Giessenbachstrasse 1, 85748 Garching, Germany
            \and
            INAF - IAPS, via Fosso del Cavaliere, 100, I-00133 Roma, Italy
              }

\date{Accepted 16 September 2022. Received 29 July 2022}

\titlerunning{CHEMOUT III. Nitrogen isotopic ratios in the outer Galaxy}
\authorrunning{Colzi et al.}

 
\abstract{Nitrogen isotopic ratios are a key tool for tracing Galactic stellar nucleosynthesis.}
{We present the first study of the \nratio\;abundance ratio in the outer regions of the Milky Way (namely, for galactocentric distances, $R_{\rm GC}$, from 12 kpc up to 19 kpc), with the aim to study the stellar nucleosynthesis effects in the global Galactic trend.}
{We analysed IRAM 30m observations towards a sample of 35 sources in the context of the CHEMical complexity in star-forming regions of the OUTer Galaxy (CHEMOUT) project. We derived the \nratio\;ratios from HCN and HNC for 14 and 3 sources, respectively, using the $J$ = 1--0 rotational transition of HN$^{13}$C, H$^{15}$NC, H$^{13}$CN, and HC$^{15}$N.}
{The results found in the outer Galaxy have been combined with previous measurements obtained in the inner Galaxy. We find an overall linear decreasing H$^{13}$CN/HC$^{15}$N ratio with increasing $R_{\rm GC}$. This translates to a parabolic \nratio\;ratio with a peak at 11~kpc. Updated Galactic chemical evolution models have been taken into account and compared with the observations. The parabolic trend of the \nratio\;ratio with $R_{\rm GC}$ can be naturally explained (i) by a model that assumes novae as the main $^{15}$N producers on long timescales ($\ge$1 Gyr) and (ii) by updated stellar yields for low- and intermediate-mass stars.}
{}


   \keywords{Galaxy: abundances -- Galaxy: evolution -- (Galaxy:) local interstellar matter -- ISM: abundances --ISM: molecules -- radio lines: ISM}

   \maketitle
   
%

\section{Introduction}
\label{intro}

Isotopic abundance ratios of carbon, nitrogen, and oxygen (CNO elements) are commonly used to trace the chemical history from molecular clouds to planetary systems (e.g. \citealt{caselliceccarelli2012}), as well as chemical enrichment and nucleosynthesis processes in galaxies (e.g. \citealt{romano2017}).
Nitrogen is the fifth most abundant element in the universe, and its isotopic ratio (\nratio) is found to be 441$\pm$6 for the proto-solar nebula in which our Sun was born (\citealt{marty2010}). This value is higher than those measured in pristine Solar System material, such as comets (144$\pm$3; \citealt{hily-blant2017}), and in carbonaceous chondrites of meteorites (44--264; \citealt{vankooten2017}). Thus, there was an enrichment of $^{15}$N during the formation of the Solar System, the causes of which are still unknown.
Observations of different molecular clouds and star-forming regions show a spread in the \nratio\;ratios depending on the physical conditions and/or molecular species. \nratio\;ratios of $\sim$150--800 have been found towards low-mass pre-stellar cores and protostellar objects, infrared dark clouds (IRDCs), and high-mass star-forming regions (e.g. \citealt{daniel2013}; \citealt{hily-blant2013a,hily-blant2013b}; \citealt{wampfler2014}; \citealt{zeng2017}; \citealt{colzi18a,colzi18b}). Moreover, \citet{guzman2017} find an HCN/HC$^{15}$N ratio of 80--160 towards a sample of protoplanetary discs, similar to the value found in comets. Finally, \nratio\;ratios from N$_{2}$H$^{+}$ towards pre-stellar cores and some massive star-forming regions behave completely differently and are higher than 800 (e.g. \citealt{fontani2015b}; \citealt{redaelli2018}).

In general, the \nratio\;ratio is known to be governed by local chemical processes and by stellar nucleosynthesis. As a local process, low-temperature isotopic-exchange reactions or different rate coefficients for isotope-substitute gas-phase reactions have been invoked to explain the observed values (e.g. \citealt{roueff2015}; \citealt{wirstrom2018}; \citealt{loison2019}; \citealt{hily-blant2020}). 
However, recent observational works have highlighted the importance of isotope-selective photodissociation of N$_{2}$ in explaining the local variation in the \nratio\;ratios in massive molecular clouds, low-mass Class 0/I objects, IRDCs, and protoplanetary discs (\citealt{colzi2019}; \citealt{bergner2020}; \citealt{fontani2020,fontani2021}; \citealt{hily-blant2021}; \citealt{evans2022}; \citealt{spezzano2022}), as suggested by chemical models (e.g. \citealt{furuya2018}; \citealt{visser2018}; \citealt{lee2021}).

Besides chemistry, the \nratio\;elemental ratio is governed by stellar nucleosynthesis, since the two elements are synthesised through different processes.
Both $^{14}$N and $^{15}$N are produced by fast-rotating massive stars as primary elements at low metallicities (\citealt{meynet2002,limongi2018}). Primary $^{14}$N is also synthesised at the base of the convective envelope of asymptotic giant branch (AGB) stars (e.g. \citealt{renzini1981}; \citealt{izzard2004}). 
Most of the $^{14}$N production from intermediate-mass stars, though, is secondary due to cold CNO processing during the main sequence and in the H-burning shells of red giants at relatively high metallicities (e.g. \citealt{karakas2014}). $^{15}$N is likely mainly produced by novae on a Galactic scale (\citealt{matteucci1991}; \citealt{romano2003}; \citealt{romano2017}), but a contribution to its production from massive stars triggered by proton ingestion in the He shell cannot be ruled out (\citealt{pignatari2015}).

Galactic chemical evolution (GCE) models that include up-to-date stellar yields predict that the $^{14}$N/$^{15}$N ratio across the disc first increases in the galactocentric distance range $R_{\rm GC}$=4--8~kpc and then stays constant or even mildly decreases up to 16~kpc (\citealt{romano2019}). The predictions from these models are confirmed by observations towards massive star-forming regions in the inner Galaxy ($R_{\rm GC}\le$12 kpc; e.g. \citealt{adande2012}; \citealt{colzi18b}), but no observations had been available to constrain GCE models to the outermost part of the Galaxy ($R_{\rm GC}>$12 kpc).

In this work we present the first study of the \nratio\;ratio towards the outer Galaxy in the context of the CHEMical complexity in star-forming regions of the OUTer Galaxy (CHEMOUT) project (\citealt{fontani2022a}, hereafter Paper I).  The observations are performed towards a sample of 35 dense molecular clouds associated with IRAS colours typical of star-forming regions, clearly detected in H$_{2}$CO $J_{K_{a}, K_{b}}$=2$_{1,2}$--1$_{1,1}$ (\citealt{blair2008}), and between 8.7 and 23.4 kpc from the Galactic centre. More information on the sources, such as their coordinates, heliocentric distances, and molecules detected, can be found in Paper I. Moreover, \citet{fontani2022b} recently studied CH$_{3}$OH, H$_{2}$CO, and HCO emission towards 15 out of the 35 targets of the CHEMOUT sample (hereafter Paper II).
In Sects.~\ref{observations} and \ref{results} we present the observations, the analysis, and the observational results. In Sect.~\ref{sec:GCE}, updated GCE models are presented. Finally, a discussion of the observational results, the comparison with GCE model predictions, and the conclusions from this work are given in Sect.~\ref{discussion}.

\section{Observations}
\label{observations}

This work is based on the observations done for the CHEMOUT project, described in Paper I, performed with the Institut de RadioAstronomie Millim\'etrique (IRAM) 30m telescope.
In the analysis presented here, we have used the 3 mm observations that include the $J= 1-0$ transitions of H$^{15}$NC, HN$^{13}$C, H$^{13}$CN, and HC$^{15}$N.
The observations were done with the Fast Fourier Transform Spectrometer (FTS) in the finest frequency resolution of 50~kHz, providing a velocity resolution of $\sim0.17$~km s$^{-1}$ at 88 GHz.
For this work all the spectra were smoothed to 0.34 km s$^{-1}$.
The data were obtained with the wobbler-switching technique with a wobbler throw of 240\asec, which translates to physical sizes of 10$-$30 pc for the distances considered in this work (8.7$-$23.4 kpc), significantly larger than the expected molecular emission of the targeted species. Other details (e.g.~pointing and focus, full spectral windows, telescope efficiencies, and weather conditions) are given in Paper I. The spectra have been converted from antenna temperature to main beam temperature ($T_{\rm MB}$; see Table 2 of Paper I). The noise achieved, $\sigma$, at the frequencies of the observed transitions is given in Appendix \ref{app-fit} for each source.

\section{Analysis and observational results}
\label{results}

\begin{table*}
\setlength{\tabcolsep}{5.5pt}
\begin{center}
\caption{Total column densities,  H$^{13}$CN/HC$^{15}$N and HN$^{13}$C/H$^{15}$NC ratios, \nratio\;ratios, and galactocentric distances ($R_{\rm GC}$).}
\begin{tabular}{lccccccc}
\hline
Source  & $N$(HN$^{13}$C)       & $N$(H$^{15}$NC) & $N$(H$^{13}$CN) & $N$(HC$^{15}$N)&   H$^{13}$CN/HC$^{15}$N\tablefootmark{a}  & HCN/HC$^{15}$N\tablefootmark{a} & $R_{\rm GC}$ \\
& ($\times$10$^{11}$ cm$^{-2}$) & ($\times$10$^{11}$ cm$^{-2}$)  & ($\times$10$^{11}$ cm$^{-2}$) &  ($\times$10$^{11}$ cm$^{-2}$) & (HN$^{13}$C/H$^{15}$NC)& (HNC/H$^{15}$NC)& (kpc) \\
\hline
WB89-315    &  $\le$1.3   &  $\le$1.1  & $\le$1.0  & $\le$1.0    &   -- & -- & 16.3 \\
WB89-379    &  1.0$\pm$0.2 &  $\le$0.5   & 6.2$\pm$0.3 & 1.5$\pm$0.2  &  4.1$\pm$0.6 ($\ge$2.1) & 452$\pm$77 ($\ge$232) & 16.4 \\
WB89-380    &  1.6$\pm$0.3\tablefootmark{b} &  $\le$0.9   & 8.2$\pm$0.5 & 2.0$\pm$0.3  &  4.1$\pm$0.7 ($\ge$1.7) & 447$\pm$90 ($\ge$189) & 16.0 \\
WB89-391    &  $\le$0.3   &  $\le$0.3   & 5.0$\pm$0.2 & 1.0$\pm$0.1  &  5.0$\pm$0.6 & 546$\pm$78 & 16.1 \\
WB89-399    &  $\le$1.1   &  $\le$1.3   & $\le$1.2 & $\le$1.2    &  -- & -- & 16.0 \\
WB89-437    &  2.2$\pm$0.4 &  $\le$1.0   & 15.5$\pm$0.6 & 4.5$\pm$0.4  &  3.5$\pm$0.3 ($\ge$2.1) & 370$\pm$48 ($\ge$226) & 15.7 \\
WB89-440    &  $\le$0.6   &  $\le$0.8   & $\le$0.7   & $\le$0.7    &  -- & -- & 15.7 \\
WB89-501    &  $\le$0.5   &  $\le$0.5   & 2.3$\pm$0.3 & 1.1$\pm$0.4\tablefootmark{b}  &  2.1$\pm$0.9 & 228$\pm$101 & 15.6 \\
WB89-529    &  $\le$1.2 &  $\le$1.6   & $\le$1.3   & $\le$1.4    &  -- & -- & 17.8 \\
WB89-572     & $\le$0.5   &  $\le$0.6   & 1.7$\pm$0.5\tablefootmark{b} & 1.0$\pm$0.2\tablefootmark{b}  &  1.6$\pm$0.6 & 196$\pm$73 & 18.3 \\
WB89-621    &  1.4$\pm$0.2 &  $\le$0.5   & 18.1$\pm$0.5 & 3.2$\pm$0.3  &  5.6$\pm$0.6 ($\ge$2.9) & 703$\pm$86 ($\ge$365) & 18.9 \\
WB89-640     & $\le$0.8   &  $\le$0.8   & 2.3$\pm$0.5\tablefootmark{b} & $\le$0.8    &  $\ge$2.8   & $\ge$312   & 16.8 \\
WB89-670    &  1.0$\pm$0.2\tablefootmark{b} &  $\le$0.7   & $\le$0.6   & $\le$0.6    &  -- ($\ge$1.4) & -- ($\ge$214) & 23.4 \\
WB89-705    &  2.9$\pm$0.3 &  $\le$0.8   & $\le$0.6   & $\le$0.6    &  -- ($\ge$3.77) & -- ($\ge$510) & 20.5 \\
WB89-789    &  2.7$\pm$0.3 &  $\le$0.8   & 3.4$\pm$0.5\tablefootmark{b} & 3.4$\pm$0.5  &  1.0$\pm$0.2 ($\ge$3.52) & 127$\pm$30 ($\ge$445) & 19.1 \\
WB89-793    &  1.9$\pm$0.7\tablefootmark{b} &  $\le$1.6   & $\le$1.5   & $\le$0.9    &  -- ($\ge$1.2) & -- ($\ge$137) & 16.9 \\
WB89-898    &  $\le$1.0   &  $\le$1.2   & 1.5$\pm$0.8\tablefootmark{b} & $\le$0.9    &  $\ge$1.6   & $\ge$173   & 15.8 \\
19423+2541  &  1.3$\pm$0.2 &  1.3$\pm$0.4\tablefootmark{b} & 14.5$\pm$0.5 & 2.7$\pm$0.5  &  5.3$\pm$0.9 (1.0$\pm$0.4) & 492$\pm$99 (96$\pm$34) & 13.5 \\
19383+2711  &  1.0$\pm$0.2 &  $\le$0.5   & 7.6$\pm$0.6 & 2.9$\pm$0.4  &  2.6$\pm$0.4 ($\ge$1.9) & 228$\pm$45 ($\ge$173) & 12.7 \\
19383+2711-b\tablefootmark{c}  & $\le$0.5   &  $\le$0.5   & 5.2$\pm$0.6 & $\le$1.3    &  $\ge$4.1   & $\ge$373   & 13.2 \\
19489+3030  &  1.9$\pm$0.2 &  $\le$0.7   & 3.5$\pm$0.5 & $\le$0.7    &  $\ge$5.0 ($\ge$2.8)   & $\ge$446 ($\ge$254)   & 12.9 \\
19571+3113  &  1.3$\pm$0.3\tablefootmark{b} &  $\le$0.9  & 4.0$\pm$0.5 & 1.1$\pm$0.2  &  3.6$\pm$0.9 ($\ge$1.5) & 311$\pm$85 ($\ge$127) & 12.2 \\
20243+3853  &  1.3$\pm$0.3\tablefootmark{b} &  $\le$0.7   & 6.0$\pm$0.6 & 1.1$\pm$0.2  &  5.5$\pm$1.2 ($\ge$1.8) & 494$\pm$119 ($\ge$159) & 12.8 \\
WB89-002  &  $\le$1.2   &  $\le$1.6   & $\le$1.3   & $\le$1.3    &  -- & -- & 8.7 \\
WB89-006  &  2.7$\pm$0.5 &  2.7$\pm$0.6 & 2.7$\pm$0.7\tablefootmark{b} & $\le$1.1   &  $\ge$2.5 (1.0$\pm$0.3)   & $\ge$246 (96$\pm$28)   & 14.3 \\
WB89-014  &  $\le$0.9   &  $\le$1.1   & $\le$0.9   & $\le$1.0    &  -- & -- & 14.9 \\
WB89-031  &  $\le$0.7   &  $\le$0.8  & $\le$0.7   & $\le$0.7    &  -- & -- & 14.1 \\
WB89-035  &  $\le$0.6   &  $\le$7.8   & 3.3$\pm$0.5 & $\le$0.6    &  $\ge$5.4  & $\ge$492   & 13.1 \\
WB89-040  &  0.60$\pm$0.16\tablefootmark{b} &  $\le$0.5   & 2.6$\pm$0.4 & $\le$0.5   &  $\ge$5.4 ($\ge$1.1)   & $\ge$454 ($\ge$97)   & 11.9 \\
WB89-060  &  2.0$\pm$0.3 &  $\le$0.8   & 18.8$\pm$0.7 & 3.0$\pm$0.3  &  6.3$\pm$0.7 ($\ge$2.4) & 595$\pm$87 ($\ge$230) & 13.6 \\
WB89-076  &  3.31$\pm$0.18 &  1.2$\pm$0.4\tablefootmark{b} & 3.3$\pm$0.4 & 1.07$\pm$0.19  &  3.1$\pm$0.6 (2.7$\pm$0.8) & 318$\pm$71 (275$\pm$87) & 15.1 \\
WB89-080  &  $\le$1.2  &  $\le$1.5   & $\le$1.3   & $\le$1.2    &  -- & -- & 12.8 \\
WB89-083  &  0.70$\pm$0.08 &  $\le$0.4   & $\le$0.4   & $\le$0.4   &  -- ($\ge$1.8) & -- ($\ge$182) & 14.7 \\
WB89-152  &  $\le$1.0   &  $\le$1.1   & $\le$1.1   & $\le$1.1    &  -- & -- & 14.4 \\
WB89-283  &  $\le$0.3 &  $\le$0.4   & 1.1$\pm$0.3\tablefootmark{b} & $\le$0.4    &  $\ge$2.6   & $\ge$278   & 15.8 \\
WB89-288  &  $\le$0.6   &  $\le$0.4   & $\le$0.4   & 0.6$\pm$0.2\tablefootmark{b}    &  $\leq$0.6 & $\leq$71 & 16.8 \\
 \hline
  \normalsize
  \label{fit-colden-ratios}
   \end{tabular}
   \end{center}
\tablefoot{Column densities have been derived assuming a $T_{\rm ex}$ of 25 K (see Sect.~\ref{sec-linean}). Column density errors do not take the calibration error ($\sim$10\%) into account. The calibration error for the \nratio\;ratio largely cancels out as the corresponding pairs of lines are recorded with the same spectral setup. \tablefoottext{a}{Ratios derived from HNC are also indicated in parentheses, when derived.}
\tablefoottext{b}{Tentative detection.}
\tablefoottext{c}{Second velocity component at $\varv_{\rm LSR}$ = -71.2 km s$^{-1}$, derived from the H$^{13}$CN(1--0) line peak (see Table ~\ref{table-h13cn}).}} 
 \end{table*}

\subsection{Detection information}

The spectroscopic information of H$^{13}$CN, HN$^{13}$C, and HC$^{15}$N is taken from the Cologne Database for Molecular
Spectroscopy\footnote{http://cdms.astro.uni-koeln.de/classic/} (CDMS; \citealt{muller2001,muller2005}; \citealt{endres2016}), and that of H$^{15}$NC from the Jet Propulsion Laboratory catalogue\footnote{https://spec.jpl.nasa.gov/ftp/pub/catalog/catdir.
html} (JPL; \citealt{pickett1998}).
 The entry of H$^{13}$CN, based on the laboratory works of \citet{fuchs2004}, \citet{cazzoli2005b}, and \citet{maiwald2000}, includes the hyperfine structure due to the $^{14}$N nucleus.
The entry of HN$^{13}$C is based on \citet{vandertak2009}; the entry of HC$^{15}$N is based on \citet{fuchs2004} and \citet{cazzoli2005a}; and the entry of H$^{15}$NC is based on \citet{creswell1976} and \citet{pearson1976}.
The rest frequencies of the $J$=1--0 transitions are: 86.3387 GHz ($F$=1--1), 86.3402 GHz ($F$=2--1), and 86.3423 GHz ($F$=0--1) for H$^{13}$CN, 87.0908 GHz for HN$^{13}$C, 86.0550 GHz for  HC$^{15}$N, and 88.8657 GHz for H$^{15}$NC. 
More spectroscopic information can be found in \citet{colzi18b}.

The observed spectra at the rest frequencies of HN$^{13}$C(1--0), H$^{15}$NC(1--0), H$^{13}$CN(1--0), and HC$^{15}$N(1--0), towards the 35 sources, are shown in Figs.~\ref{fig-spectra-hn13c-1}--\ref{fig-spectra-hc15n-1} (Appendix \ref{sec-spectra}). 
The three H$^{13}$CN(1--0) hyperfine components, which are separated by 5 and 7 km s$^{-1}$, can be resolved towards all sources since the linewidths are narrower (0.7$-$3.6 km s$^{-1}$; see Sect. \ref{res-results}).

We consider a line detected when the integrated intensity signal-to-noise ratio (S/N)  is $>$6. Moreover, we consider a line tentatively detected when the integrated intensity S/N is between 4 and 6.
H$^{13}$CN(1--0) has been detected in 21 sources (60\% of the total sample). The three hyperfine components are clearly detected towards six sources: WB89-379, WB89-391, WB89-437, WB89-621, 19423+2541, and WB89-060 (see Fig.~\ref{fig-spectra-h13cn-1}). Among the H$^{13}$CN detections, HC$^{15}$N(1--0) has been detected in 14 sources (40\% of the total sample).
HN$^{13}$C(1--0) has been detected in 18 sources (51\% of the total sample), and H$^{15}$NC(1--0) only in 3 of them ($\sim$9\% of the total sample).

The profile of the HN$^{13}$C(1--0) transition, which is the only species detected towards WB89-670, shows an inverse P-Cygni profile, and this might suggest infall material towards the centre of the source. For visualisation purposes, we have fitted this line with two velocity components, one in emission and one in absorption assuming a continuum background temperature of 50 K. 
A similar profile was also observed in Paper I in HCO$^{+}$ and c-C$_{3}$H$_{2}$ towards the same source.
As also found in Paper I for c-C$_{3}$H$_{2}$, towards 19383+2711 we detect two velocity components that are probably part of the same cloud. Conversely, for 19571+3113  we only detect the velocity component at $-$61 km s$^{-1}$ and not that at $-$66 km s$^{-1}$ (see Paper I).

\subsection{Molecular line analysis}
\label{sec-linean}

\begin{figure*}
\centering
\includegraphics[width=21pc]{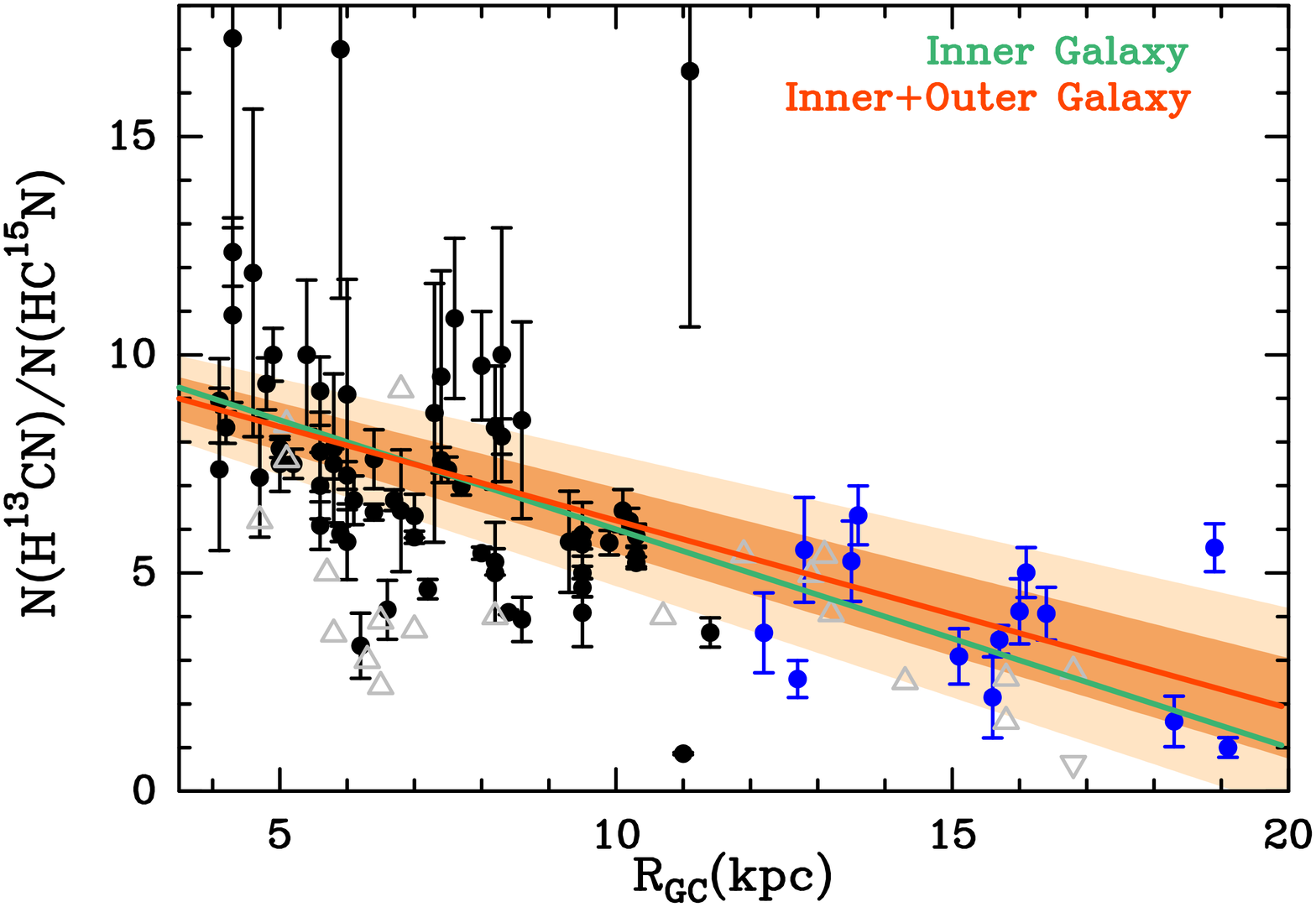}
\includegraphics[width=21pc]{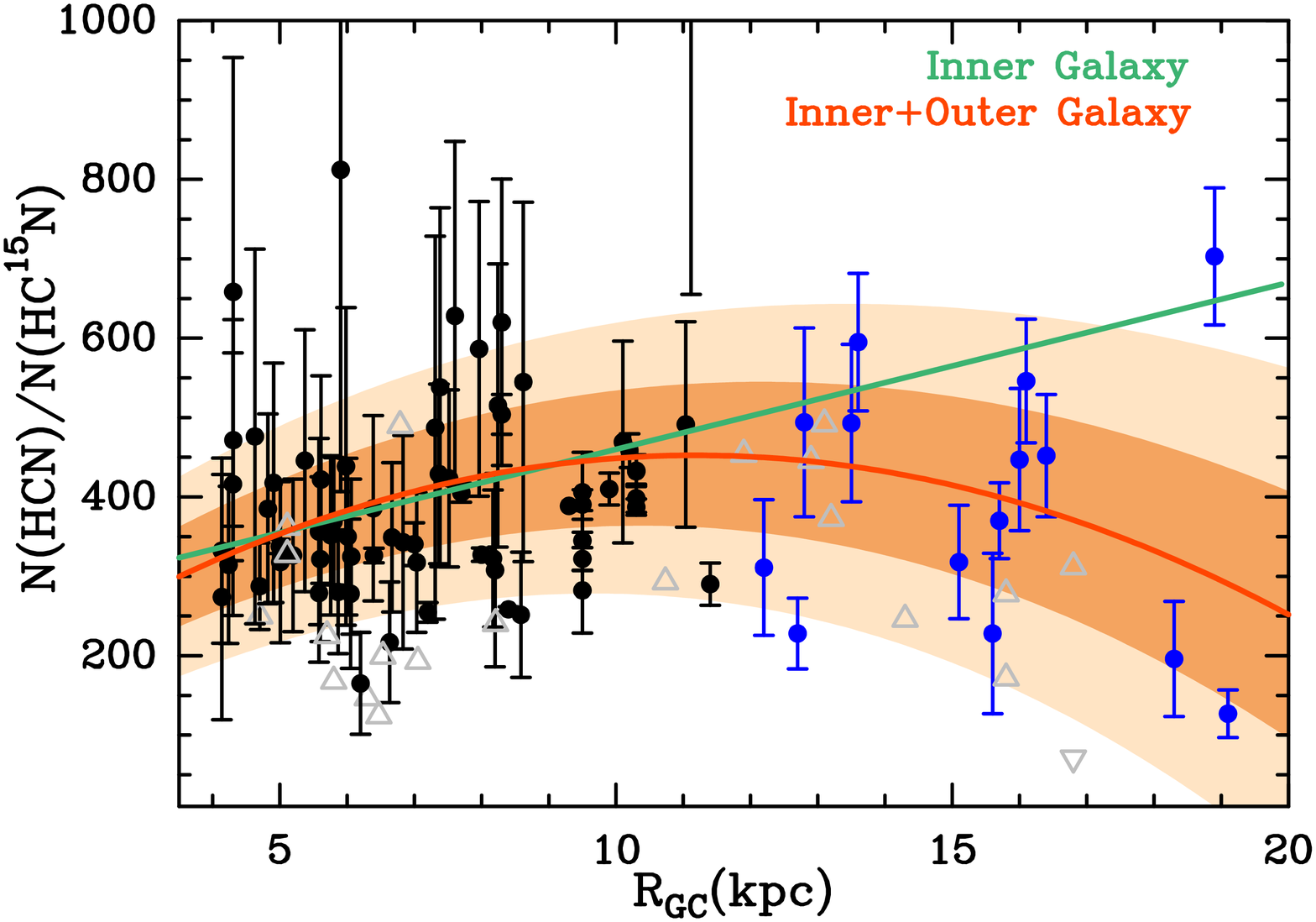}
\caption{Isotopic ratios vs. galactocentric distance. \emph{Left panel:} H$^{13}$CN/HC$^{15}$N ratio as a function of the galactocentric distance, $R_{\rm GC}$. The solid red line is the linear regression fit obtained using the whole sample, along with its 0.5$\sigma$ and 1$\sigma$ uncertainty (dark and light orange regions, respectively). \emph{Right panel:} Same as the left panel but for the \nratio\;ratio of HCN. The solid red parabola is obtained from the analysis of the data, as explained in Sect.~\ref{res-results}, along with its 0.5$\sigma$ and 1$\sigma$ uncertainty (dark and light orange regions, respectively). In both panels the solid green line represents the linear trend found in the inner Galaxy by \citet{colzi18b} and extrapolated up to 20 kpc. Black points represent the values obtained in the inner Galaxy by \citet{colzi18b}, and blue points are the values obtained in this work towards the outer Galaxy. Grey triangles pointing upwards are lower limits, while those pointing downwards are upper limits. Note that the source at $R_{\rm GC}$=2.1 kpc from \citet{colzi18b} has been used for the analysis but is not shown in this plot for visualisation purposes.}
\label{fig-ratio-without68}
\end{figure*}

First, for each line we fitted the baseline with a polynomial function of maximum order 1. Then we used the Spectral Line Identification and Modeling (SLIM) tool within the MADCUBA package\footnote{Madrid Data Cube Analysis on ImageJ is a software developed at the Center of Astrobiology (CAB) in Madrid; \url{https://cab.inta-csic.es/madcuba/}.} (\citealt{martin2019}).
SLIM generates a synthetic spectrum, assuming local thermodynamic equilibrium (LTE) conditions considering the line opacity, and applies the AUTOFIT algorithm to find the best non-linear least-square fit to the data.
The free parameters that are considered in the fit are the column density of the molecule, $N$, the excitation temperature, $T_{\rm ex}$, the velocity, $\varv_{\rm LSR}$, and the full width half maximum (FWHM; see details in \citealt{martin2019}). 

The column densities were evaluated assuming that the emission fills the telescope beam (i.e. no beam dilution has been applied) since we do not have any size measurement of the emitting region of these lines in this sample of sources. Therefore, the derived column densities are beam-averaged values.
For the analysis of H$^{13}$CN, we considered the three hyperfine components to perform the fit with SLIM.
The $T_{\rm ex}$ of the targeted molecules cannot be derived since we have a single rotational transition. We thus assumed that the $J$=1$-$0 rotational levels of
the four molecules (H$^{13}$CN, HC$^{15}$N, HN$^{13}$C, and H$^{15}$NC) are populated with the same $T_{\rm ex}$. The non-LTE calculations performed by \citet{colzi18b} on a sample of star-forming regions in the inner Galaxy showed that this is a reasonable assumption, since the $T_{\rm ex}$ of the $J$=1$-$0 transitions of the molecules differ only within $\sim$16$\%$ (see their Sect. 3.2).
For the fit procedure we assumed a $T_{\rm ex}$ of 25 K for all the sources analysed, and hence the column density derived (see Table \ref{fit-colden-ratios}) corresponds to this temperature. However, we note that while the column densities of the individual species depend on $T_{\rm ex}$, their ratios, which are the main goal of the work, are almost independent of it. As discussed by \citet{colzi18a}, the \nratio\;derived from the $J$=1$-$0 rotational transition does not depend significantly on the assumed $T_{\rm ex}$. 
The analysis of other molecules for which multiple transitions have been detected towards this sample (CH$_{3}$CCH and CH$_{3}$OH) have provided values of $T_{\rm ex}$=7$-$27 K (Paper I and II, respectively).  These species can be used as thermometers and are expected to trace an extended molecular envelope similar to that traced by HCN.
Using this range as a guide, we explored how the assumption of $T_{\rm ex}$ affects the derived molecular column density ratios. By changing the assumed $T_{\rm ex}$ between 5 K and 30 K, we find that the H$^{13}$CN/HC$^{15}$N ratio only varies between 1\% and 10\% with respect to the values obtained with $T_{\rm ex}$ = 25 K.  
The SLIM fit also provides an estimate of the line opacities and confirms that the transitions of all the isotopologues are optically thin.
Upper limits on column densities were also derived for undetected lines. They were determined taking the 3$\sigma$ root mean square of the spectra at the observed rest frequencies into account (see Appendix \ref{app-fit}) and assuming the FWHM. The latter was considered to be equal to that of the other corresponding isotopologue (e.g. that of H$^{13}$CN for the upper limit of HC$^{15}$N), if detected, or to that of the other molecules (e.g. that of HN$^{13}$C or H$^{15}$NC, or the average of the two, for the upper limit of HC$^{15}$N) for the same source. If none of the four molecules have been detected, we assumed the same FWHM as that of c-C$_{3}$H$_{2}$, which is detected in all of the sources, except WB89--315 for which we used the FWHM of HCO$^{+}$. 
Moreover, for upper limits, the $\varv_{\rm LSR}$ was assumed to be equal to the velocity, $\varv_{0}$, obtained from H$_{2}$CO observations by \citet{blair2008} (see Paper I and Fig.~\ref{fig-spectra-hn13c-1}).

\subsection{Results}
\label{res-results}

The results from the fit procedure are listed in Tables \ref{fit-hn13c}, \ref{fit-h15nc}, \ref{fit-h13cn}, and \ref{fit-hc15n} for HN$^{13}$C(1--0), H$^{15}$NC(1--0), H$^{13}$CN(1--0), and HC$^{15}$N(1--0), respectively. Total column densities, H$^{13}$CN/HC$^{15}$N and HN$^{13}$C/H$^{15}$NC ratios, and $R_{\rm GC}$ are listed in Table \ref{fit-colden-ratios}. We derived the \nratio\;ratios by multiplying H$^{13}$CN/HC$^{15}$N and HN$^{13}$C/H$^{15}$NC by \cratio\;as a function of the $R_{\rm GC}$ derived by \citet{milam2005} for CN, which is a nitrile species, as are HCN and HNC:
\begin{equation}
^{12}\textrm{C}/^{13}\textrm{C}=(6.01\pm1.19) \textrm{ kpc}^{-1}\times R_{\rm GC} +(12.28\pm9.33).
\label{milam}
\end{equation}
Other similar galactocentric trends, which are consistent within the associated uncertainties, have also been derived using H$_2$CO and CO (\citealt{milam2005}; \citealt{yan2019}). Since in this work we analyse nitriles and isonitriles, we adopted the trend obtained from CN shown in Eq.~\ref{milam}.
This relation was obtained for the inner Galaxy, so its extrapolation to the outer Galaxy is uncertain. To our best knowledge there are no observational studies of the behaviour of the \cratio\;ratio with $R_{\rm GC}$ in the outer Galaxy towards a large sample of sources. \citet{wouterloot1996} derived the $^{13}$CO/C$^{18}$O ratio towards only five sources (WB89-380, WB89-391, WB89-399, WB89-437, and WB89-501, whose $R_{\rm GC}$ are about 16 kpc; see Table \ref{fit-colden-ratios}), and they found $^{13}$CO/C$^{18}$O ratios in the range 12--17.5. 
These values can be converted to the \cratio\;ratio assuming a value of $^{16}$O/$^{18}$O. Although this ratio has not been studied observationally in the outer Galaxy, GCE models (\citealt{romano2017}) predict values of $\sim$2000. This leads to a \cratio\;ratio for CO of 115--170, significantly higher than the local value of $\sim$70. Moreover, \citet{wouterloot1996} directly derived a \cratio\;ratio of 200$\pm$15 towards WB89-437 from C$^{18}$O/$^{13}$C$^{18}$O. \citet{milam2005} also studied the \cratio\;ratio towards WB89-391 and found a value of 134$\pm$43 for CN, in very good agreement with that of CO ($\sim$132) obtained by \citet{wouterloot1996}. These estimates suggest that the \cratio\;ratio keeps increasing in the outer Galaxy, in good agreement with the extrapolation of the galactocentric trend of Eq.~\eqref{milam}. We stress that this is also supported by the predictions from GCE models, as described in more detail in Sect.~\ref{sec:GCE} and Appendix \ref{sec:c-ratio-GCE}. Hence, we have assumed that the linear trend of Eq. \ref{milam} is also valid for the outer Galaxy, as we did for the inner Galaxy in \citet{colzi18b}.

The uncertainties of the \nratio\;ratios were evaluated by considering the uncertainty given by the fit procedure and propagating it by also taking Eq.~\eqref{milam} into account.  The final \nratio\;ratios obtained are listed in Table \ref{fit-colden-ratios}. 
Since H$^{15}$NC has only been detected in three sources, the galactocentric trends and the comparison with GCE models are discussed just for HCN from now on.

The left panel of Fig.~\ref{fig-ratio-without68} shows the H$^{13}$CN/HC$^{15}$N ratio as a function of $R_{\rm GC}$ for the sources with $R_{\rm GC}$ from 12 kpc to 19 kpc derived in this work, and for those with $R_{\rm GC}$ between 2 kpc and 12 kpc from \citet{colzi18b}. 
For the entire set of data we performed an unweighted linear regression fit and find
\begin{equation}
\label{eq-1}
\textrm{H}^{13}\textrm{CN}/\textrm{HC}^{15}\textrm{N}=(-0.43\pm0.08) \textrm{ kpc}^{-1}\times R_{\rm GC} +(10.5\pm0.7).
\end{equation}
This fit is very similar to the extrapolation of the linear trend found by \citet{colzi18b} in the inner Galaxy (see the red line with respect to the green line in the left panel of Fig.~\ref{fig-ratio-without68}).

The right panel of Fig.~\ref{fig-ratio-without68} shows the \nratio\;ratios of HCN as a function of $R_{\rm GC}$. 
Most of the observed points in the outer Galaxy stay clearly below the extrapolation up to 20 kpc of the linear \nratio\;trend obtained in the inner Galaxy by \citet{colzi18b}.
This indicates that the increasing \nratio\;ratio with $R_{\rm GC}$ found in the inner Galaxy is not valid beyond 10-12 kpc. Indeed, the observed \nratio\;values in the outer Galaxy decrease with $R_{\rm GC}$, with the only exception being WB89-621, which presents a \nratio\;ratio of $\sim$700 at 18.9 kpc. This source is one of the most luminous and massive of the sample (see Paper I and \citealt{elia2021}), suggesting the presence of a protocluster whose chemistry could affect the \nratio\;ratio. However, higher-angular-resolution observations are needed to draw conclusions.

Following the analysis done by \citet{colzi18b} for the inner Galaxy, and multiplying Eq.~\eqref{eq-1} by Eq.~\eqref{milam}, a parabolic trend is obtained (see the red curve in the right panel of Fig.~\ref{fig-ratio-without68}, where the 0.5$\sigma$ and 1$\sigma$ uncertainties are also shown), with a maximum at $\sim$11 kpc:
\begin{equation}
    \label{eq-2}
    {\rm HCN}/{\rm HC}^{15}{\rm N}= -2.58 \textrm{ kpc}^{-2}\times R_{\rm GC}^{2} +57.82 \textrm{ kpc}^{-1}\times R_{\rm GC} + 128.94.
\end{equation}

In Appendix \ref{sec-azimutalvar} we also look for azimuthal \nratio\;ratio variations as already done in the inner Galaxy by \citet{colzi18b}. No trend within the spiral arms is found (see Fig.~\ref{fig-azimuthal-vari}).

\section{Galactic chemical evolution models}
\label{sec:GCE}

\begin{figure*}
\centering
\includegraphics[width=21pc]{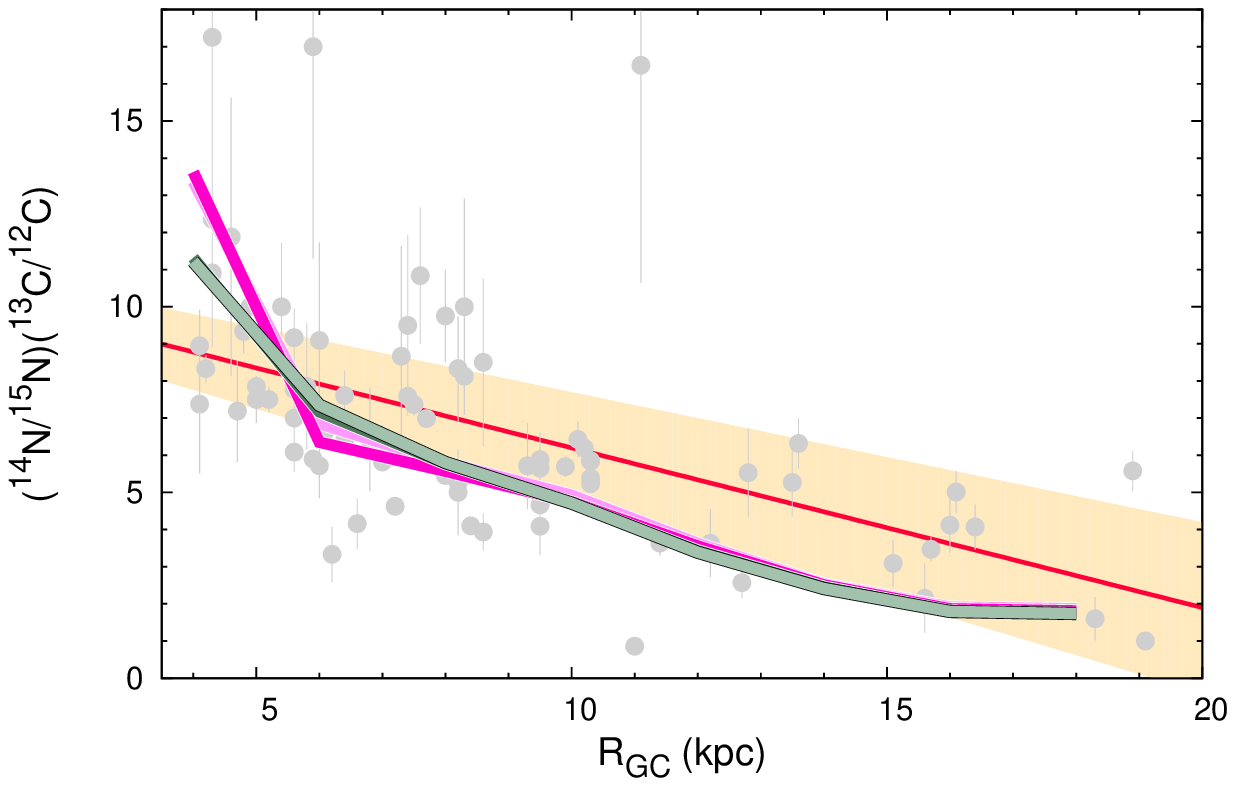}
\includegraphics[width=21pc]{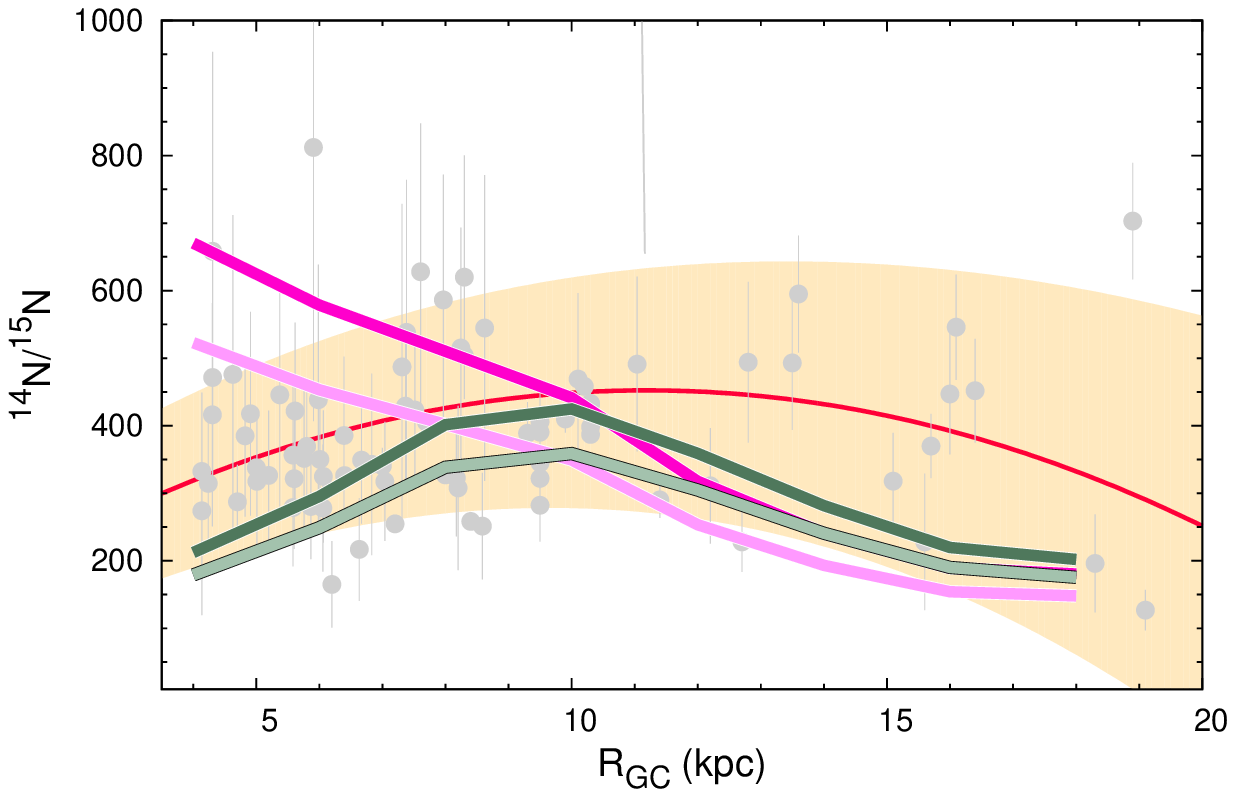}
\caption{Isotopic ratios predicted by GCE models. \emph{Left panel:} Comparison between the ($^{14}$N/$^{15}$N)$\times$($^{13}$C/$^{12}$C) ratios derived from observations and predicted by GCE models. The red line and light orange regions are the same as in Fig.~\ref{fig-ratio-without68}. \emph{Right panel:} Same as the left panel, but for the $^{14}$N/$^{15}$N ratio. In both panels grey points correspond to the observed star-forming regions. The dark green, light green, magenta, and pink lines are the predictions from GCE models (see Sect.~\ref{sec:GCE} and Table \ref{table-models}). Note that in the left panel the dark green model is below the light green one.}
\label{grad_mod1}
\end{figure*}

To understand the galactocentric trend observed, we studied the evolution of the chemical composition of the interstellar medium in the Galactic disc at different $R_{\mathrm{GC}}$ using the chemical evolution model described in \cite{grisoni2017,grisoni2018} and \cite{romano2021}. According to this model, the Milky Way disc forms inside-out (\citealt{matteucci1989}) with a higher star-formation efficiency in the inner regions, such that a metallicity gradient is naturally established, as observed (e.g. \citealt{mendez2022}).

The adopted nucleosynthesis prescriptions follow \cite{romano2019,romano2021}, with some updates (Romano et al., in prep.). For single low- and intermediate-mass stars ($1 \le M/{\mathrm M}_\odot < 9$), we adopted the yields of \cite{ventura2013,ventura2014,ventura2018,ventura2020,ventura2021} that cover all metallicity regimes, from the ultra metal-poor to the super-solar. The adopted yields include the effects of dredge ups, hot bottom burning, and mass loss, as well as a proper treatment of the super-AGB phase for the most massive stars. The yields of single massive stars ($13 \le M/{\mathrm M}_\odot \le 100$) that end their lives as core-collapse supernovae are taken from \cite{limongi2018}. In particular, for [Fe/H]~$< -1$~dex we used their `set R' (see \citealt{limongi2018} for details), assuming that all stars have initial rotation velocities $\varv_{\mathrm{rot}} =$~300 km~s$^{-1}$. For higher metallicities ([Fe/H]~$\ge -1$) we assumed the yields for non-rotating stars. The mass limit for full collapse to black holes is set to 30~M$_\odot$ if [Fe/H]~$< -1$ and to 60~M$_\odot$ if [Fe/H]~$\ge -1$. These choices allow us to reproduce satisfactorily well the evolution of the stable CNO isotopes in the Galaxy (\citealt{romano2019}; Romano et al., in prep.).

The model implements $^{13}$C and $^{15}$N production from nova systems following \cite{romano2017,romano2021}. First, the rate of formation of nova systems at any time is computed as a fraction, $\alpha$, of the white dwarf (WD) birth rate at a previous time, taking into account the delay needed for the WDs to cool to a temperature that ensures strong enough nova outbursts  ($\sim$1 Gyr). The free parameter $\alpha$ is assumed to be constant in time, and its value was set as to reproduce the current nova rate in the Galaxy of $R_{\mathrm{nova}} (t_{\mathrm{now}}) \simeq$ 35 yr$^{-1}$ (\citealt{de2021}). In computing the theoretical nova rate, it was assumed that each nova suffers $10^4$ outbursts during its lifetime (\citealt{bath1978}); for the sake of simplicity, the outbursts were assumed to occur instantaneously. The average masses ejected in the form of $^{13}$C and $^{15}$N by each nova were fixed by the request that the relevant CNO isotope observations be reproduced; they are listed in Table~\ref{table-models}. We caution that there is a high degeneracy in the suggested nova yields due to the uncertain current Galactic nova rate (e.g. \citealt{de2021}; \citealt{rector2022}).

The four models presented in this work are summarised in Table \ref{table-models}.  We ran models in which the nova system WD progenitors have initial masses in the range 1--8~M$_\odot$ and models in which only the most massive WDs (originating from stars in the range 3--8 M$_{\odot}$) have the right characteristics to produce nova outbursts (see \citealt{romano2021}). The predicted gradients are compared to the observational trends in Fig.~\ref{grad_mod1}. 
We further considered different values of the masses ejected in the form of $^{13}$C and $^{15}$N in each outburst (see Table \ref{table-models}). In fact, rather than providing a `best-fitting curve', we aimed to show a `permitted area'. We stress that the nova yields listed in Table \ref{table-models} are linked to the assumed current nova rate (35 yr$^{-1}$). Should the rate be sensibly higher or lower than this (e.g. \citealt{de2021}; \citealt{rector2022}), the yields would have to be changed accordingly.
For the sake of completeness, we also present the predicted $^{12}$C/$^{13}$C gradient in Appendix \ref{sec:c-ratio-GCE}, which is in good agreement with the observational trend we have used in this work, from \citet{milam2005}.

\begin{table}
\begin{center}
\caption{\label{table-models}GCE model prescriptions.}
\begin{tabular}{lcccc}
\hline
Model   & Mass range for & M$_{\mathrm{ejec}}^{\mathrm{^{13}C}}$ & M$_{\mathrm{ejec}}^{\mathrm{^{15}N}}$ & Flag \\
            & WD\tablefootmark{a} progenitors   & & \\
            & (M$_\odot$) & (M$_\odot$) & (M$_\odot$) & \\
\hline
1 & 1--8 & 5.40e$-$7 & 3.35e$-$8 & dark green \\
2 & 1--8 & 6.40e$-$7 & 3.95e$-$8 & light green \\
3 & 3--8 & 3.25e$-$7 & 1.91e$-$8 & magenta \\
4 & 3--8 & 4.25e$-$7 & 2.40e$-$8 & pink \\
\hline
\end{tabular}
\tablefoot{Col.~2: mass range of primary stars in nova systems; Cols.~3 and 4: average ejected masses of $^{13}$C and $^{15}$N per nova outburst.
\tablefoottext{a}{White dwarf.}}
\end{center}
\end{table}

\section{Discussion and conclusions}
\label{discussion}

In this work we have studied, for the first time, the \nratio\;ratio towards 35 star-forming regions located in the outer Galaxy ($R_{\rm GC}>$12 kpc). We have derived the H$^{13}$CN/HC$^{15}$N ratio towards 14 of them and the HN$^{13}$C/H$^{15}$NC ratio for 3 sources. 
We have observed a clearly decreasing trend of the H$^{13}$CN/HC$^{15}$N ratio with increasing $R_{\rm GC}$ [Eq.~\eqref{eq-1}], also taking previous observations of the inner Galaxy by \cite{colzi18b} into account. This decreasing trend has a very similar slope as that found previously only in the inner Galaxy.
Extrapolating the linear positive \cratio\;trend derived by \citealt{milam2005} to the outer Galaxy, which is supported by GCE models, the \nratio\;ratio shows a parabolic trend increasing up to 11 kpc and then decreasing for larger galactocentric distances [Eq.~\eqref{eq-2}].

If $^{15}$N is produced during nova outbursts on long timescales ($\ge$1 Gyr), as assumed in the adopted GCE models\footnote{Novae are also necessary to explain $^7$Li enrichment in the Galaxy (e.g. \citealt{romano2021}, and references therein).}, the observed trend of increasing $^{14}$N/$^{15}$N when moving from the inner Galaxy to the solar circle, and decreasing $^{14}$N/$^{15}$N when moving from the solar radius to the outer disc, can be reproduced. In particular, if low-mass stars ($M < 1.5$ M$_\odot$) enter the formation of nova systems, significant $^{15}$N pollution is expected in the inner Galaxy from $M \sim$1 M$_\odot$ stars that formed in large number in the early Milky Way, due to the very efficient star-formation rates and faster formation of these regions. In the outer Galaxy, which suffers low-level star formation and slow gas accretion from the intergalactic medium (inside-out formation), the effect is reduced. In fact, the differences between the case in which low- and intermediate-mass stars enter the formation of nova systems with respect to that in which this occurs only for intermediate-mass stars are minimal (cf. magenta and pink versus green curves in Fig. \ref{grad_mod1}). The decrease in the $^{14}$N/$^{15}$N ratio in the outer Galaxy, on the other hand, is dictated by the nucleosynthesis prescriptions for single low- and intermediate-mass stars: \cite{romano2019} show that a decreasing trend has to be expected in the outer Galaxy when adopting the stellar yields of \cite{ventura2013,ventura2014,ventura2018,ventura2020,ventura2021} due to the strong metal dependence of the $^{14}$N yield in this case. Their predictions are confirmed by the observations presented in this work. 
As shown in Appendix \ref{sec:c-ratio-GCE}, GCE simulations predict that the \cratio\;ratio also keeps increasing with $R_{\rm GC}$  in the outer Galaxy, independently of the assumed model (Fig.~\ref{grad_mod3} in Appendix \ref{sec:c-ratio-GCE}). This is consistent with the assumption we made {a priori} to evaluate the \nratio\;ratios (Sect.~\ref{res-results}) from the observed H$^{13}$CN/HC$^{15}$N ratios. However, this theoretical prediction needs to be confirmed with more observations of the \cratio\;ratio in the outer Galaxy.

While the general galactocentric behaviour of the \nratio\;ratio can be mainly explained by nucleosynthesis effects, its scatter at each galactocentric distance could be associated with local chemical fractionation effects, such as isotope selective photodissociation of N$_{2}$ (e.g. \citealt{furuya2018}), as mentioned in Sect.~\ref{intro}. We are not able to discuss this point in more detail since physical properties, such as H$_{2}$ densities, kinetic temperatures, and the possible presence of protostellar objects, are not available -- or not well constrained -- for all objects at present. It is clear from these new observations in the outer Galaxy that nitrogen fractionation effects, if present, are not systematic or dominant in the outer Galaxy, as already found in the inner Galaxy. This is true for the spatial scales studied in these works ($\sim$0.2--2 pc). In fact, the average trend with $R_{\rm GC}$ is consistent -- within the associated uncertainties -- with the predictions of GCE models that take stellar nucleosynthesis effects into account but do not account for chemical processes acting in molecular clouds. Higher-angular-resolution observations would be needed for each source to disentangle the local contribution from the nucleosynthetic one. 

\begin{acknowledgements}
We thank the anonymous referee for the careful reading of the article and the useful comments.
This work is based on observations carried out under projects number 116-17 and 004-18 with the IRAM 30m telescope. IRAM is supported by INSU/CNRS (France), MPG (Germany) and IGN (Spain).
F.F. is grateful to the IRAM 30m staff for their precious help during the observations.
L.C. acknowledges financial support through the Spanish grant PID2019-105552RB-C41 funded by MCIN/AEI/10.13039/501100011033.
V.M.R. has received support from the Comunidad de Madrid through the Atracción de Talento Investigador Modalidad 1 (Doctores con experiencia) Grant (COOL:Cosmic Origins of Life; 2019-T1/TIC-5379), and the Ayuda RYC2020-029387-I funded by MCIN/AEI /10.13039/501100011033.
This publication has received funding from the European Union Horizon 2020 research and innovation programme under grant agreement No 730562 (RadioNet).
\end{acknowledgements}

\bibliographystyle{aa} 
 \bibliography{bibliography} 
%
%

%


\begin{appendix} 

\section{Fit results}
\label{app-fit}

In this appendix the results of the fitting procedure for the HN$^{13}$C(1--0), H$^{15}$NC(1--0), H$^{13}$CN(1--0), and HC$^{15}$N(1--0) lines of all sources
are shown. A $T_{\rm ex}$ of 25 K has been assumed for all the sources and molecules. The line analysis is explained in Sect.~\ref{sec-linean}. 

\begin{table}[h!]
\setlength{\tabcolsep}{2pt}
\begin{center}
\caption{Results from the LTE fitting procedure described in Sect.~\ref{sec-linean} for HN$^{13}$C(1--0).}
\begin{tabular}{lcccc}
\hline
Source  & $\varv_{\rm LSR}$     & FWHM & $\int T_{\rm MB}d\nu$ & $\sigma$\\
& (km s$^{-1}$) & (km s$^{-1}$)   & (mK km s$^{-1}$) &  (mK) \\
\hline
WB89-315  & -95.1 &  2.1 & $\le$32 & 13.0 \\
WB89-379 & -89.1$\pm$0.2 &  2.2$\pm$0.5 & 33$\pm$13 & 5.8 \\
WB89-380 & -85.7$\pm$0.3 &   3.0$\pm$0.7 & 50$\pm$20\tablefootmark{a} & 7.0 \\
WB89-391 & -85.71 &  1.1 & $\le$11 & 6.0 \\
WB89-399  & -82.2 &  2.2 & $\le$36 & 14.0 \\
WB89-437  & -71.4$\pm$0.3 &  3.2$\pm$0.6 & 70$\pm$23 & 7.4 \\
WB89-440  & -72.2 &  1.6 & $\le$20 & 9.0 \\
WB89-501 & -58.4 &  1.3 & $\le$16 & 8.0 \\
WB89-529 & -60.1 &  1.9 & $\le$39 & 11.0 \\
WB89-572  & -48.0 &  0.7 & $\le$17 & 12.0 \\
WB89-621  & -25.40$\pm$0.07 &  0.98$\pm$0.16 & 45$\pm$13 & 6.6 \\
WB89-640  & -25.4 &  0.9 & $\le$27 & 16.0 \\
WB89-670  & -17.72$\pm$0.06 &  0.55$\pm$0.15 & 32$\pm$13\tablefootmark{a} & 14.0 \\
WB89-705  & -12.16$\pm$0.05 &  0.86$\pm$0.11 & 91$\pm$21 & 12.0 \\
WB89-789 & 34.39$\pm$0.08 &  1.39$\pm$0.19 & 87$\pm$20 & 10.0 \\
WB89-793  & 30.0$\pm$0.3 &  1.7$\pm$0.7 & 61$\pm$44\tablefootmark{a} & 16.0 \\
WB89-898  & 63.4 &  2.3 & $\le$28 & 11.0 \\
19423+2541  & -71.9$\pm$0.2 &  2.2$\pm$0.5 & 42$\pm$16 & 6.0 \\
19383+2711  & -65.74$\pm$0.18 &  1.6$\pm$0.4 & 31$\pm$15 & 6.5 \\
19383+2711-b  & -71.2 &  1.6 & $\le$16 & 6.5 \\
19489+3030 & -69.18$\pm$0.07 &  1.11$\pm$0.17 & 60$\pm$16 & 10.5 \\ 
19571+3113 & -61.0$\pm$0.3 &  2.34$\pm$0.7 & 41$\pm$19\tablefootmark{a} & 8.2 \\ 
20243+3853 & -73.5$\pm$0.4 &  2.96$\pm$0.9 & 41$\pm$20\tablefootmark{a} & 7.0 \\ 
WB89-002  & -2.8 &  1.3 & $\le$37 & 19.0 \\
WB89-006  & -91.0$\pm$0.2 &  2.4$\pm$0.5 & 86$\pm$29 & 12.4 \\
WB89-014  & -96.0 &  1.7 & $\le$28 & 12.6 \\
WB89-031  & -89.4 &  1.6 & $\le$21 & 10.0 \\
WB89-035  & -77.7 &  2.2 & $\le$19 & 7.5 \\
WB89-040  & -62.18$\pm$0.11 &  0.8$\pm$0.3 & 19$\pm$10\tablefootmark{a} & 8.0 \\ 
WB89-060  & -83.99$\pm$0.10 &  1.4$\pm$0.3 & 64$\pm$20 & 10.0 \\
WB89-076  & -96.99$\pm$0.05 &  1.7$\pm$0.11 & 104$\pm$11 & 7.0 \\
WB89-080  & -74.2 &  2.9 & $\le$39 & 13.0 \\
WB89-083  & -94.00$\pm$0.04 &  0.73$\pm$0.09 & 23$\pm$5 & 7.6 \\
WB89-152  & -88.1 &  0.8 & $\le$30 & 19.5 \\
WB89-283  & -94.5 &  1.2 & $\le$11 & 5.7 \\
WB89-288  & -100.9 &  1.6 & $\le$12 & 8.0 \\
 \hline
  \normalsize
    \label{fit-hn13c}
   \end{tabular}
   \end{center}
\tablefoot{Parameters without errors are assumed in the fitting procedure, as explained in Sect.~\ref{sec-linean}.
\tablefoottext{a}{Tentative detection.}} 
 \end{table}

\begin{table}
\setlength{\tabcolsep}{2pt}
\begin{center}
\caption{Same as Table \ref{fit-hn13c} but for H$^{15}$NC(1--0).}
\begin{tabular}{lcccc}
\hline
Source  & $\varv_{\rm LSR}$     & FWHM & $\int T_{\rm MB}d\nu$ & $\sigma$\\
& (km s$^{-1}$) & (km s$^{-1}$)   & (mK km s$^{-1}$) &  (mK) \\
\hline
WB89-315  &  -95.1 &  2.1 &  $\le$29  &  11.0  \\
WB89-379  &  -89.3 &  2.2 &  $\le$13  &  5.0  \\
WB89-380  &  -86.7 &  3.0 &  $\le$23  &  8.0  \\
WB89-391  &  -86.1 &  1.1 &  $\le$9  &  6.0  \\
WB89-399  &  -82.2 &  2.2 &  $\le$33  &  14.0  \\
WB89-437  &  -71.7 &  3.2 &  $\le$28  &  7.0  \\
WB89-440  &  -72.2 &  1.6 &  $\le$20  &  9.0  \\
WB89-501  &  -58.4 &  1.3 &  $\le$14  &  8.0  \\
WB89-529  &  -60.1 &  1.9 &  $\le$42  &  11.0  \\
WB89-572  &  -48.0 &  0.7 &  $\le$17   &  12.0  \\
WB89-621  &  -25.4 &  1.8 &  $\le$13  &  6.6  \\
WB89-640  &  -25.4 &  0.9 &  $\le$22  &  16.0  \\
WB89-670  &  -17.5 &  0.5 &  $\le$19  &  14.0  \\
WB89-705  &  -12.1 &  0.9 &  $\le$20  &  12.0  \\
WB89-789  &  34.7 &  1.4 &  $\le$20  &  10.0  \\
WB89-793  &  30.5 &  1.7 &  $\le$41  &  16.0  \\
WB89-898  &  63.4 &  2.3 &  $\le$30  &  11.0  \\
19423+2541  &  -71.3$\pm$0.4 &  2.7$\pm$0.9 &  33$\pm$18\tablefootmark{a}  &  7.0  \\ 
19383+2711   &  -65.7 &  1.6 &  $\le$13  &  6.0  \\
19383+2711-b &  -71.2 &  1.6 &  $\le$13  &  6.0  \\
19489+3030 &  -68.7 &  1.1 &  $\le$18  &  10.5  \\
19571+3113 &  -62.5 &  2.3 &  $\le$22  &  8.3  \\
20243+3853 &  -73.1 &  3.0 &  $\le$19  &  6.2  \\
WB89-002  &  -2.8 &  1.3 &  $\le$41  &  19.0  \\ 
WB89-006  &  -90.3$\pm$0.3 &  2.6$\pm$0.7 &  72$\pm$31  &  11.7  \\ 
WB89-014  &  -96.0 &  1.7 &  $\le$27  &  12.6  \\ 
WB89-031  &  -89.4 &  1.6 &  $\le$20  &  10.0  \\ 
WB89-035  &  -77.7 &  2.2 &  $\le$20  &  7.5  \\ 
WB89-040  &  -62.5 &  0.8 &  $\le$13  &  8.7  \\ 
WB89-060  &  -84.3 &  1.4 &  $\le$22  &  11.0  \\ 
WB89-076  &  -97.5$\pm$0.3 &  2.1$\pm$0.7 &  32$\pm$19\tablefootmark{a}  &  7.11  \\ 
WB89-080  &  -74.2 &  2.9 &  $\le$39  &  13.0  \\ 
WB89-083  &  -93.8 &  0.7 &  $\le$10  &  7.6  \\ 
WB89-152  &  -88.1 &  0.8 &  $\le$28  &  19.5  \\ 
WB89-283  &  -94.5 &  1.2 &  $\le$12  &  5.7  \\ 
WB89-288  &  -100.9 &  1.6 &  $\le$9  &  8.0  \\
 \hline
  \normalsize
  \label{fit-h15nc}
   \end{tabular}
   \end{center}
 \end{table}

\begin{table}
\setlength{\tabcolsep}{2pt}
\begin{center}
\caption{\label{table-h13cn}Same as Table \ref{fit-hn13c} but for H$^{13}$CN(1--0).}
\begin{tabular}{lcccc}
\hline
Source  & $\varv_{\rm LSR}$     & FWHM & $\int T_{\rm MB}d\nu$\tablefootmark{a} & $\sigma$\\
& (km s$^{-1}$) & (km s$^{-1}$)   & (mK km s$^{-1}$) &  (mK) \\
\hline
WB89-315  &  -95.1  &  2.1  &  $\le$17  & 12.0 \\
WB89-379  &  -89.20$\pm$0.04  &  1.68$\pm$0.09  &  102$\pm$11  & 5.0 \\
WB89-380  &  -86.81$\pm$0.14  &  3.5$\pm$0.3  &  136$\pm$23  & 8.0 \\
WB89-391  &  -85.98$\pm$0.03  &  1.31$\pm$0.07  &  83$\pm$10  & 6.0 \\
WB89-399  &  -82.2  &  2.2  &  $\le$20  & 14.0 \\
WB89-437  &  -71.65$\pm$0.04  &  2.3$\pm$0.1  &  255$\pm$23  & 7.0 \\
WB89-440  &  -72.2  &  1.6  &  $\le$12  & 9.0 \\
WB89-501  &  -58.37$\pm$0.06  &  0.88$\pm$0.14  &  39$\pm$13  & 8.0 \\
WB89-529  &  -60.1  &  1.9  &  $\le$21   & 11.0 \\
WB89-572  &  -47.41$\pm$0.10  &  0.7$\pm$0.3  &  28$\pm$18\tablefootmark{b}  & 12.4 \\
WB89-621  &  -25.30$\pm$0.03  &  1.83$\pm$0.06  &  299$\pm$20  & 6.6 \\
WB89-640  &  -25.10$\pm$0.09  &  0.9$\pm$0.2  &  39$\pm$18\tablefootmark{b}  & 15.0 \\
WB89-670  &  -17.5  &  0.5  &  $\le$11 & 14.0 \\
WB89-705  &  -12.1  &  0.9  &  $\le$10  & 12.0 \\
WB89-789  &  34.41$\pm$0.08  &  0.97$\pm$0.18  &  56$\pm$21  & 10.0 \\
WB89-793  &  30.4  &  1.7  &  $\le$25  & 16.0 \\
WB89-898  &  63.1$\pm$0.3  &  2.3$\pm$0.6  &  58$\pm$31\tablefootmark{b}  & 11.0 \\
19423+2541&  -72.38$\pm$0.09  &  3.61$\pm$0.16  &  242$\pm$24  & 7.7 \\
19383+2711&  -65.85$\pm$0.12  &  2.9$\pm$0.3 &  125$\pm$22  & 6.0 \\
19383+2711-b &  -71.20$\pm$0.18  &  2.9$\pm$0.4  &  87$\pm$22  & 6.0 \\
19489+3030 &  -69.12$\pm$0.11  &  1.7$\pm$0.3  &  58$\pm$18 & 10.5 \\
19571+3113 &  -62.67$\pm$0.11  &  1.8$\pm$0.3  &  66$\pm$20 & 8.3 \\
20243+3853 &  -73.30$\pm$0.12  &  2.6$\pm$0.3  &  100$\pm$22  & 6.0 \\
WB89-002  &  -2.8  &  1.3  &  $\le$21  & 19.0 \\
WB89-006  &  -91.6$\pm$0.4  &  2.7$\pm$0.7  &  45$\pm$28\tablefootmark{b}  & 9.9 \\
WB89-014  &  -96.0 &  1.7  &  $\le$15  & 12.6 \\
WB89-031  &  -89.4  &  1.6  &  $\le$11  & 10.0 \\
WB89-035  &  -77.86$\pm$0.16  &  2.2$\pm$0.4  &  54$\pm$19  & 7.6 \\
WB89-040  &  -62.40$\pm$0.07  &  0.99$\pm$0.16  &  43$\pm$14  & 8.7 \\
WB89-060  &  -84.08$\pm$0.04  &  2.15$\pm$0.09  &  312$\pm$29  & 10.0 \\
WB89-076  &  -96.87$\pm$0.07  &  1.37$\pm$0.17  &  55$\pm$14  & 7.0 \\
WB89-080  &  -74.2  &  2.9  &  $\le$22  & 13.0 \\
WB89-083  &  -93.8  &  0.7  &  $\le$6  & 7.6 \\
WB89-152  &  -88.1  &  0.8  &  $\le$18  & 19.5 \\
WB89-283  &  -94.57$\pm$0.12  &  1.2$\pm$0.3  &  18$\pm$13\tablefootmark{b}  & 6.0 \\
WB89-288  &  -100.9  &  0.5  &  $\le$7  & 8.0 \\
 \hline
  \normalsize
  \label{fit-h13cn}
   \end{tabular}
   \end{center}
\tablefoot{\tablefoottext{a}{The line integrated intensity of the main hyperfine transition ($F$=2--1) is shown.}
\tablefoottext{b}{Tentative detection.}} 
 \end{table}

\begin{table}
\setlength{\tabcolsep}{2pt}
\begin{center}
\caption{Same as Table \ref{fit-hn13c} but for HC$^{15}$N(1--0).}
\begin{tabular}{lcccc}
\hline
Source  & $\varv_{\rm LSR}$     & FWHM & $\int T_{\rm MB}d\nu$ & $\sigma$\\
& (km s$^{-1}$) & (km s$^{-1}$)   & (mK km s$^{-1}$) &  (mK) \\
\hline
WB89-315    &  -95.1  &  2.1  & $\le$31 &  12.0 \\
WB89-379    &  -89.34$\pm$0.09  &  1.4$\pm$0.2  & 45$\pm$12 &  5.0 \\
WB89-380    &  -85.3$\pm$0.4  &  3.5  & 59$\pm$24 &  7.6 \\
WB89-391    &  -85.98$\pm$0.04  &  0.9$\pm$0.1  & 30$\pm$6 &  6.0 \\
WB89-399    &  -82.2  &  2.2  & $\le$37 &  14.0 \\
WB89-437    &  -71.52$\pm$0.13  &  3.0$\pm$0.3  & 133$\pm$23 &  7.0 \\
WB89-440    &  -72.2  &  1.6  & $\le$20 &  9.0 \\
WB89-501    &  -58.6$\pm$0.4  &  1.7$\pm$0.8  & 32$\pm$26\tablefootmark{a} &  7.6 \\
WB89-529    &  -60.1  &  1.9  & $\le$4 &  11.0 \\
WB89-572     &  -48.23$\pm$0.07  &  0.60$\pm$0.15  & 32$\pm$15\tablefootmark{a} &  12.0 \\
WB89-621    &  -25.57$\pm$0.11  &  2.4$\pm$0.3  & 96$\pm$18 &  6.6 \\
WB89-640     &  -25.4  &  0.9  & $\le$24 &  16.0 \\
WB89-670    &  -17.5  &  0.5  & $\le$17 &  14.0 \\
WB89-705    &  -12.1  &  0.9  & $\le$18 &  12.0 \\
WB89-789    &  34.3$\pm$0.2  &  1.2$\pm$0.5  & 35$\pm$29\tablefootmark{a} &  10.0 \\
WB89-793    &  30.3  &  1.7  & $\le$27 &  16.0 \\
WB89-898    &  63.4  &  2.3  & $\le$28 &  11.0 \\
19423+2541  &  -72.1$\pm$0.4  &  5$\pm$1  & 81$\pm$27 &  6.4 \\
19383+2711  &  -65.90$\pm$0.14  &  2.0$\pm$0.3  & 50$\pm$14 &  6.0 \\
19383+2711-b   &  -71.2  &  2.9  & $\le$16 &  6.0 \\
19489+3030  &  -68.4  &  1.6  & $\le$21 &  10.5 \\
19571+3113  &  -62.29$\pm$0.11  &  1.1$\pm$0.3  & 33$\pm$14 &  8.3 \\
20243+3853  &  -73.20$\pm$0.15  &  1.6$\pm$0.4  & 32$\pm$13 &  7.0 \\
WB89-002  &  -2.7  &  1.3  & $\le$38 &  19.0 \\  
WB89-006  &  -91.4  &  2.7  & $\le$33 &  11.5 \\  
WB89-014  &  -96.0  &  1.7  & $\le$28 &  12.6 \\  
WB89-031  &  -89.4  &  1.6  & $\le$21 &  10.0 \\  
WB89-035  &  -77.7  &  2.2  & $\le$18 &  7.5 \\  
WB89-040  &  -62.5  &  1.0  & $\le$14 &  8.7 \\  
WB89-060  &  -84.01$\pm$0.08  &  1.7$\pm$0.2  & 88$\pm$17 &  10.0 \\  
WB89-076  &  -97.18$\pm$0.11  &  1.4$\pm$0.3  & 32$\pm$11 &  7.0 \\  
WB89-080  &  -74.2  &  2.9  & $\le$37 &  13.0 \\  
WB89-083  &  -93.8  &  0.7  & $\le$12 &  7.6 \\  
WB89-152  &  -88.1  &  0.8  & $\le$33 &  19.5 \\  
WB89-283  &  -94.5  &  1.2  & $\le$13 &  5.7 \\  
WB89-288  &  -101.16$\pm$0.07  &  0.5  & 19$\pm$12\tablefootmark{a} &  8.0 \\   
 \hline
  \normalsize
  \label{fit-hc15n}
   \end{tabular}
   \end{center}
 \end{table}

 \clearpage
\section{Spectra}
\label{sec-spectra}
In this appendix the observed spectra at the rest frequencies of HN$^{13}$C(1--0), H$^{15}$NC(1--0), H$^{13}$CN(1--0), and HC$^{15}$N(1--0) for all the sources are shown. Each spectrum is centred around the velocity, $\varv_{0}$, obtained from H$_{2}$CO observations by \citet{blair2008}. Detected and undetected transitions are indicated with red and blue curves, respectively.

\begin{figure*}[h!]
\centering
\includegraphics[width=32pc]{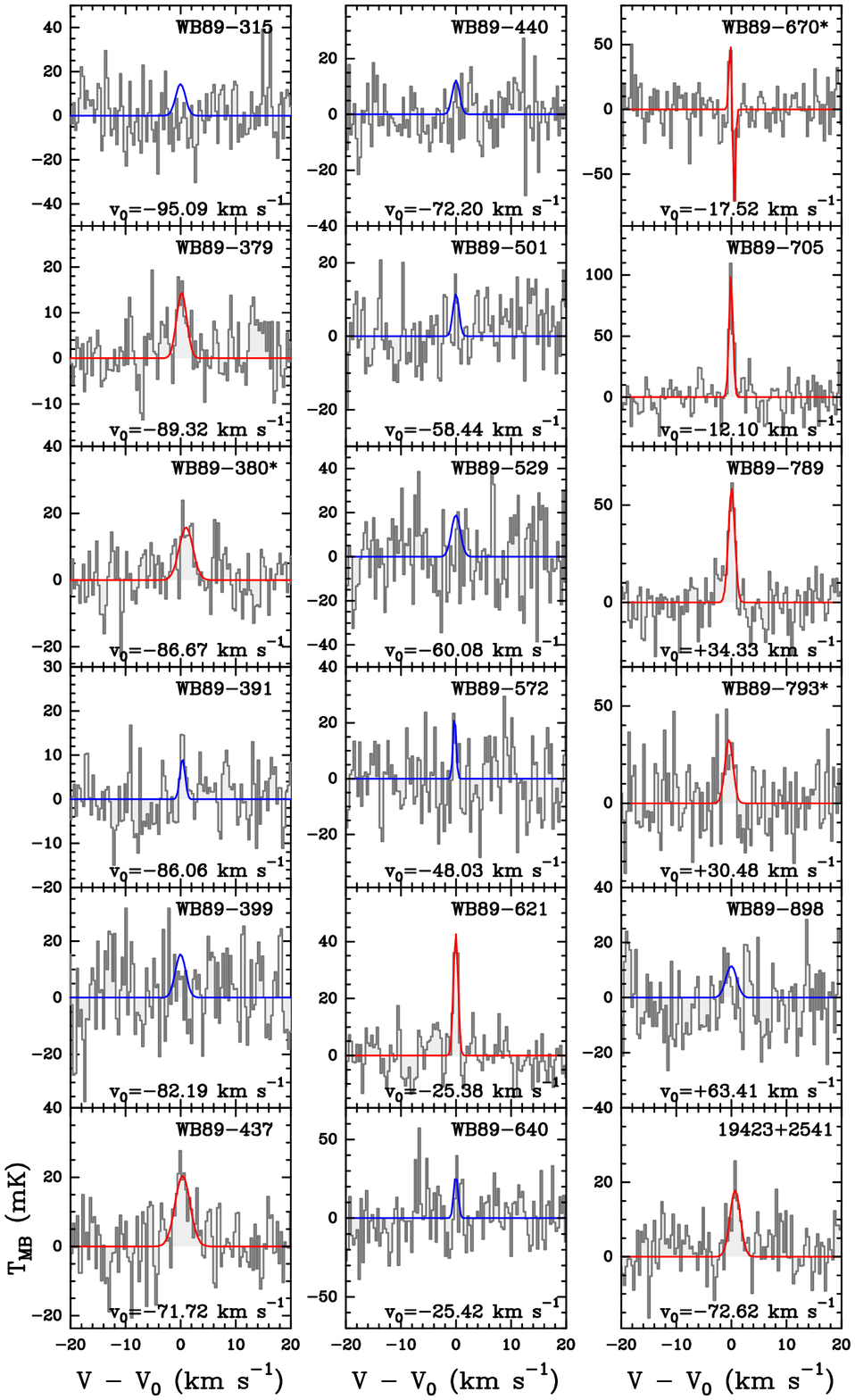}
\caption{Spectra of HN$^{13}$C(1--0) obtained for the sample of sources. For each spectrum the \emph{x} axis represents a velocity interval of $\pm$20 km s$^{-1}$ around the velocity, $\varv_{0}$, obtained from H$_{2}$CO observations by \citet{blair2008}. The \emph{y} axis shows the intensity in main beam temperature units. The red curves are the best LTE fits obtained with MADCUBA for detections and tentative detections (indicated with an asterisk in the source name). The blue curves correspond to the upper limits obtained for non-detections.}
\label{fig-spectra-hn13c-1}
\end{figure*}

\begin{figure*}[h!]
\centering
\addtocounter{figure}{-1}
\includegraphics[width=32pc]{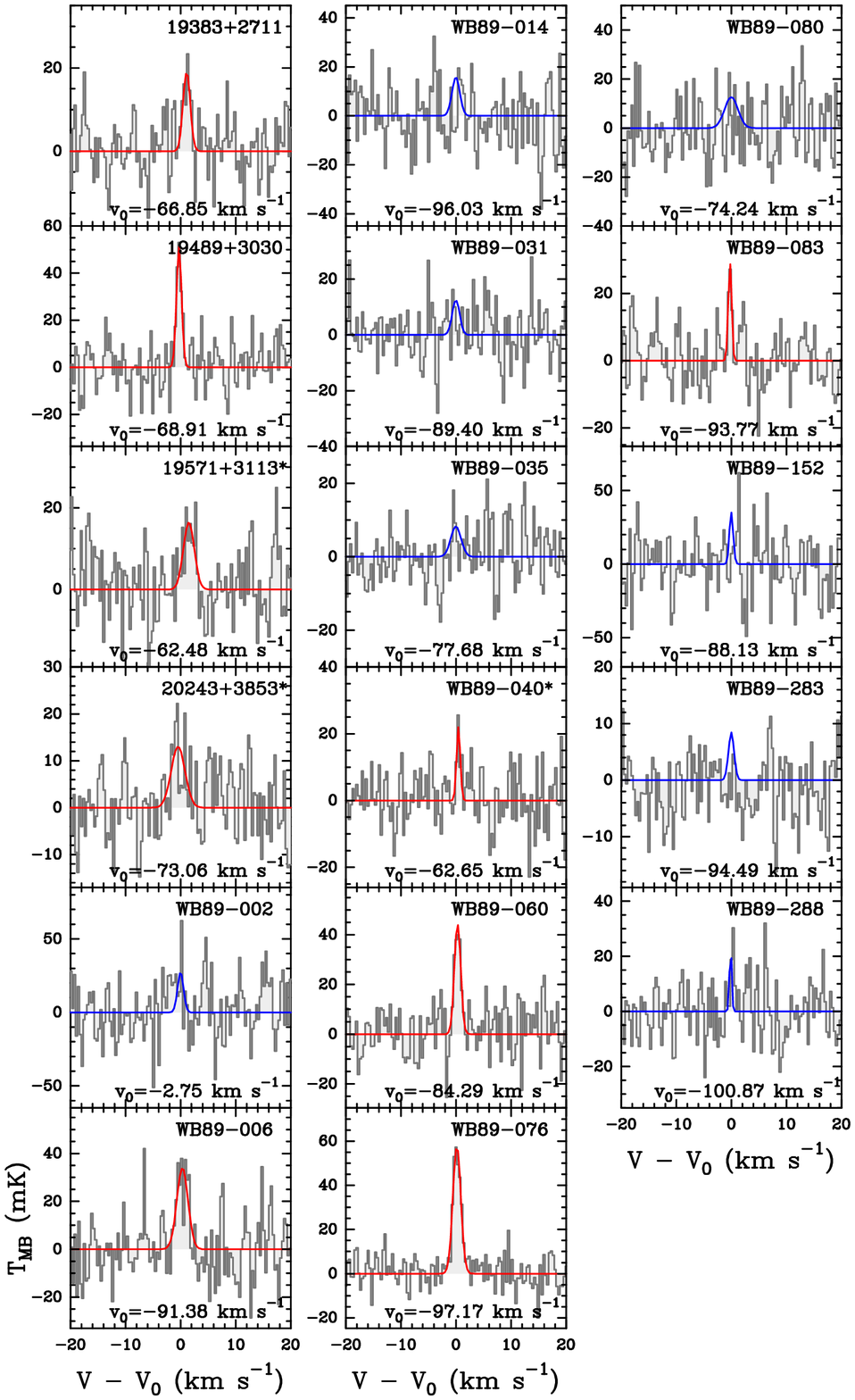}
\caption{Continued.}
\label{fig-spectra-hn13c-2}
\end{figure*}

\begin{figure*}[h!]
\centering
\includegraphics[width=32pc]{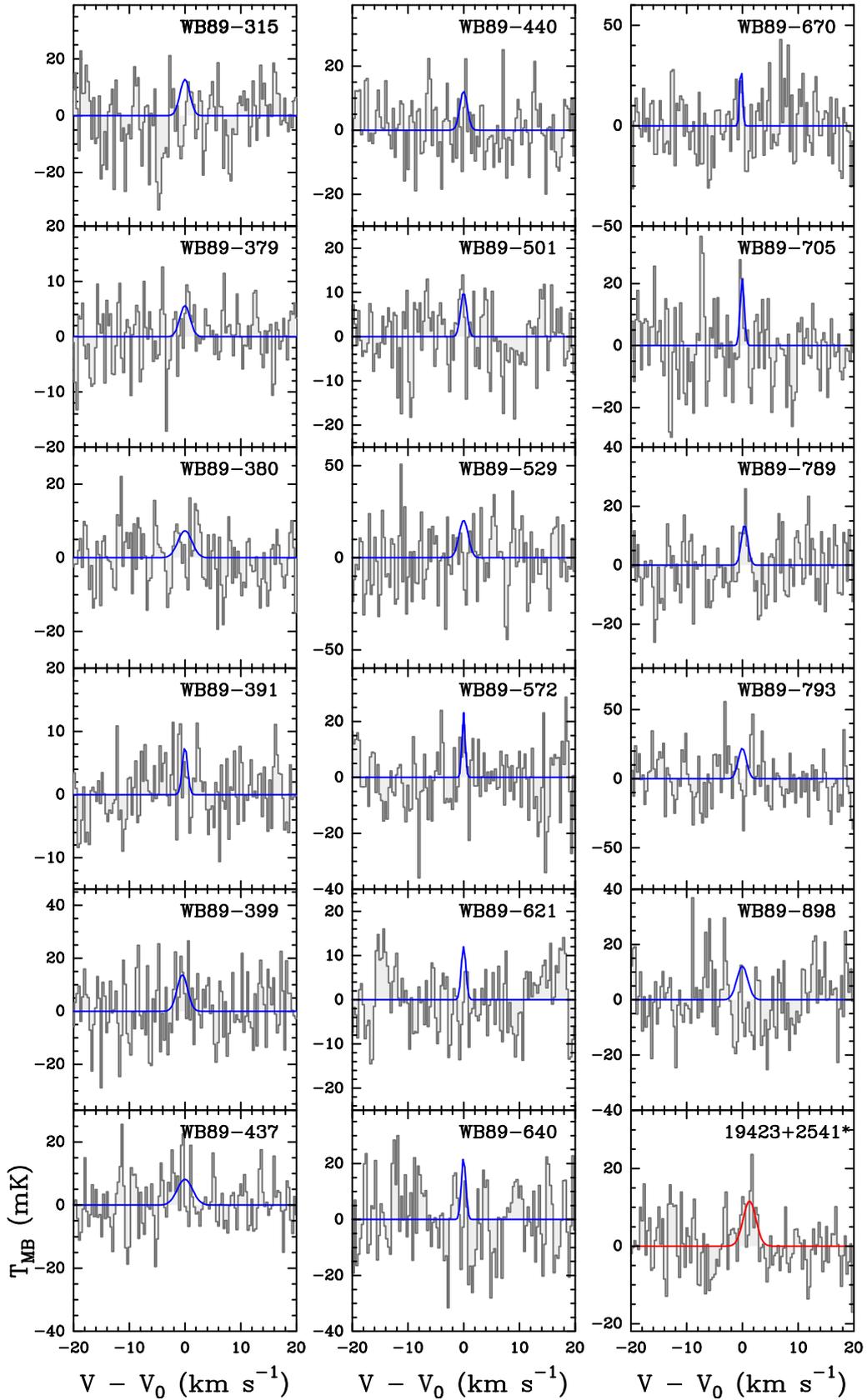}
\caption{Same as Fig.~\ref{fig-spectra-hn13c-1} but for H$^{15}$NC(1--0). The velocity, $\varv_{0}$, for each source is given in Fig.~\ref{fig-spectra-hn13c-1}.}
\label{fig-spectra-h15nc-1}
\end{figure*}

\begin{figure*}[h!]
\centering
\addtocounter{figure}{-1}
\includegraphics[width=32pc]{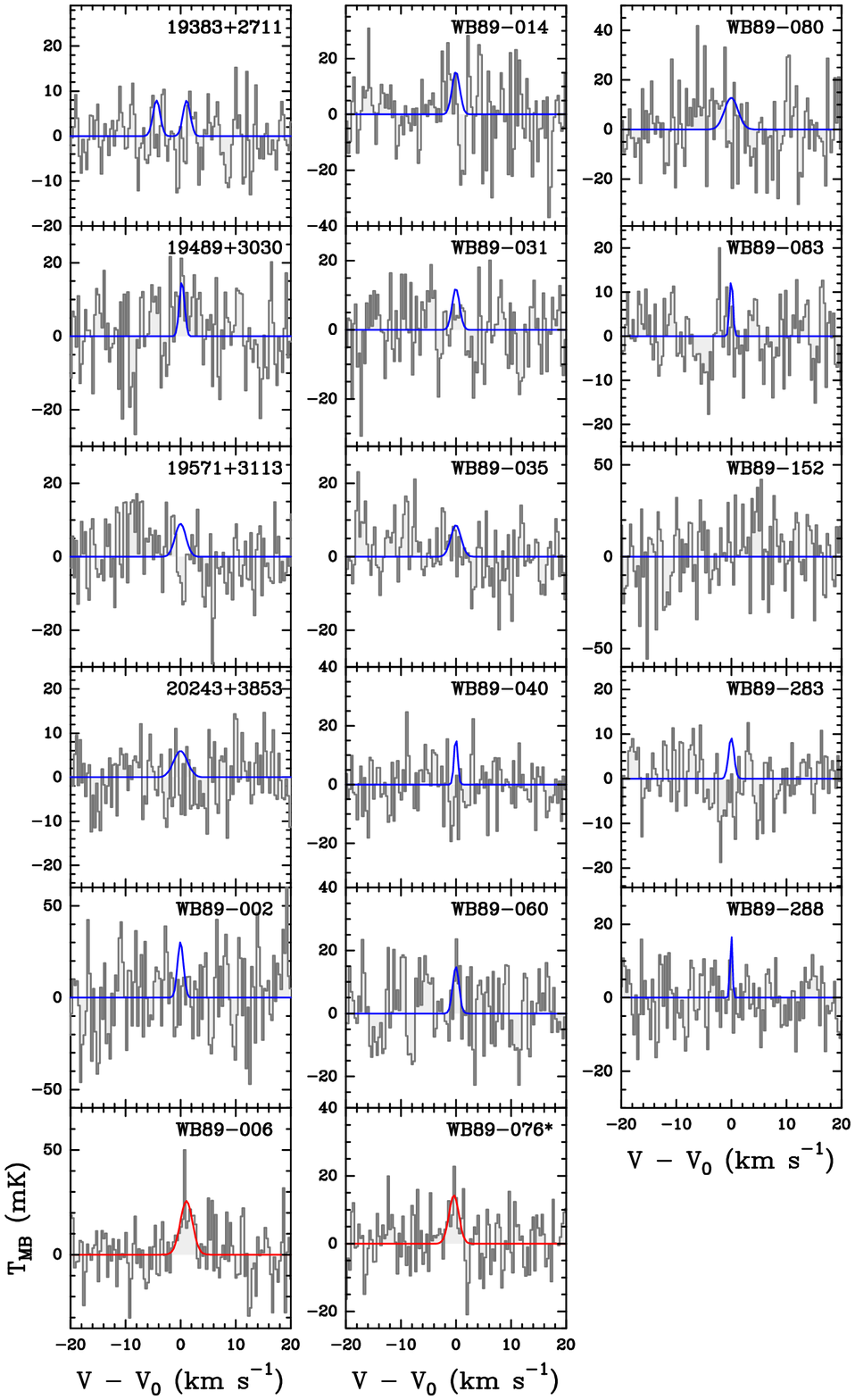}
\caption{Continued. }
\label{fig-spectra-h15nc-2}
\end{figure*}

\begin{figure*}[h!]
\centering
\includegraphics[width=32pc]{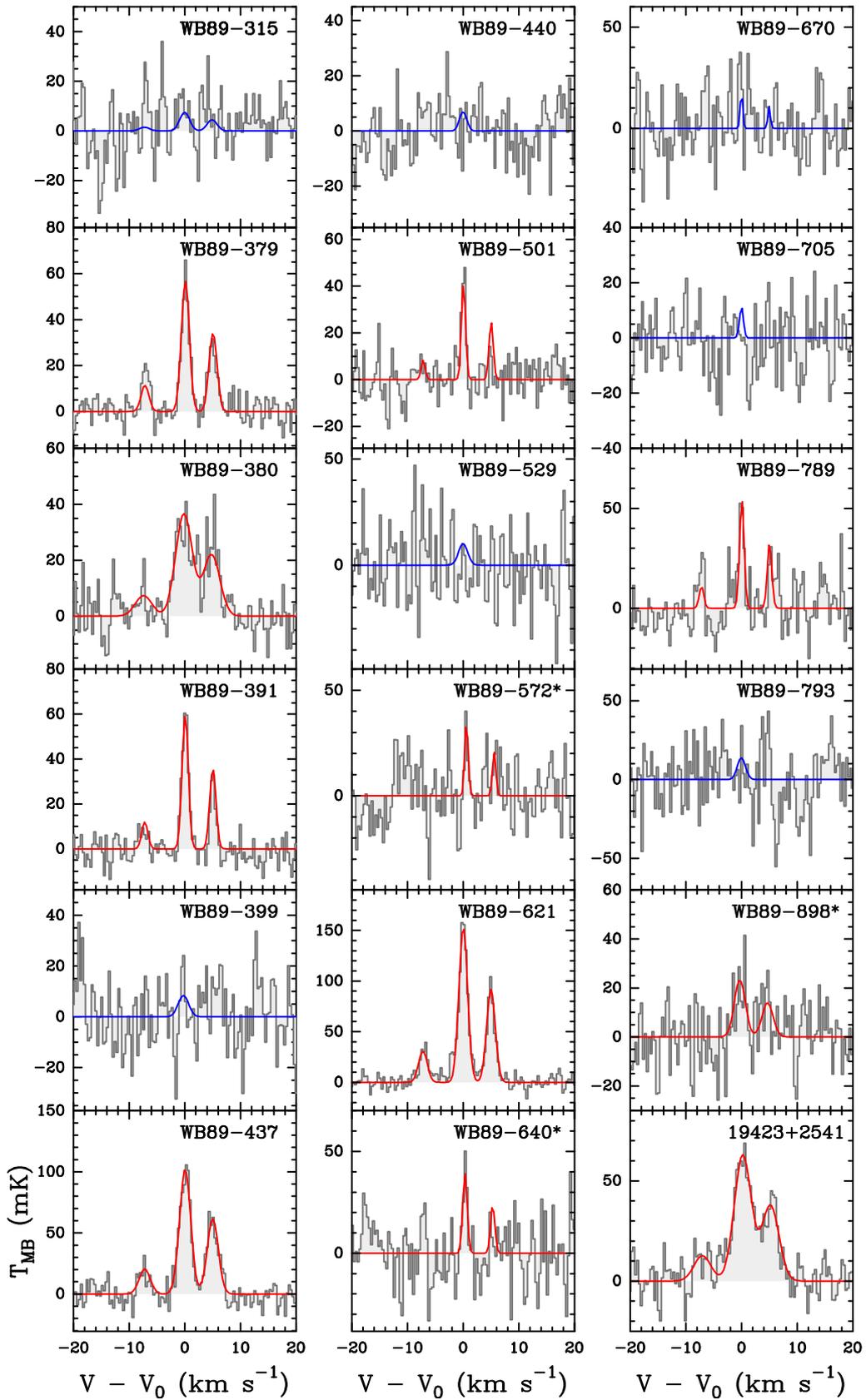}
\caption{Same as Fig.~\ref{fig-spectra-hn13c-1} but for H$^{13}$CN(1--0). The velocity, $\varv_{0}$, for each source is given in Fig.~\ref{fig-spectra-hn13c-1}. For display purposes, only the hyperfine components with LTE peak line intensities above $\sigma$/3 are shown.}
\label{fig-spectra-h13cn-1}
\end{figure*}

\begin{figure*}[h!]
\centering
\addtocounter{figure}{-1}
\includegraphics[width=32pc]{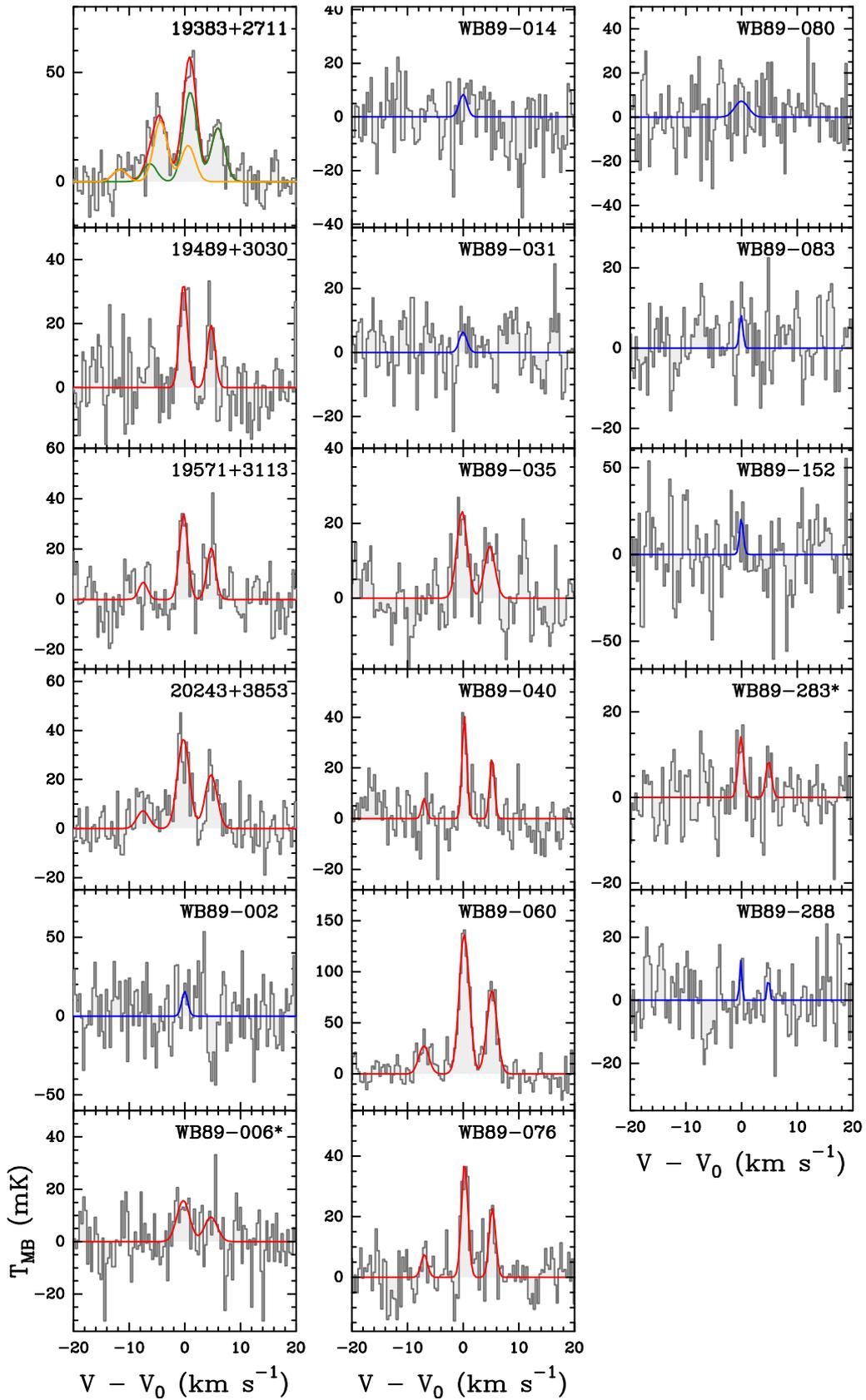}
\caption{Continued. Note that for the source 19383+2711 we have fitted two velocity components (orange curve and green curve). In this case the red curve is the sum of the LTE fit of the two components.}
\label{fig-spectra-h13cn-2}
\end{figure*}

\begin{figure*}[h!]
\centering
\includegraphics[width=32pc]{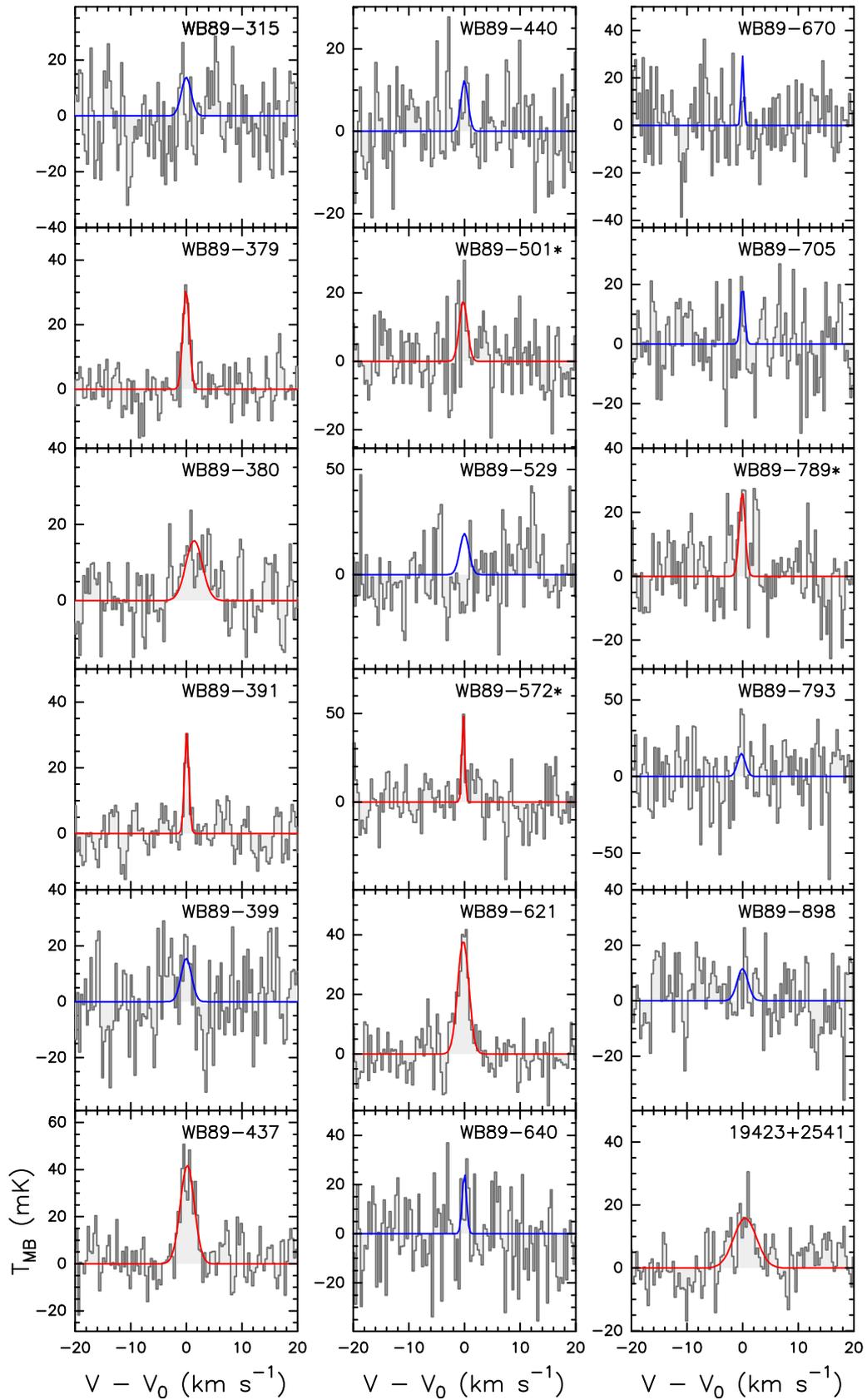}
\caption{Same as Fig.~\ref{fig-spectra-hn13c-1} but for HC$^{15}$N(1--0). The velocity, $\varv_{0}$, for each source is given in Fig.~\ref{fig-spectra-hn13c-1}.}
\label{fig-spectra-hc15n-1}
\end{figure*}

\begin{figure*}[h!]
\centering
\addtocounter{figure}{-1}
\includegraphics[width=32pc]{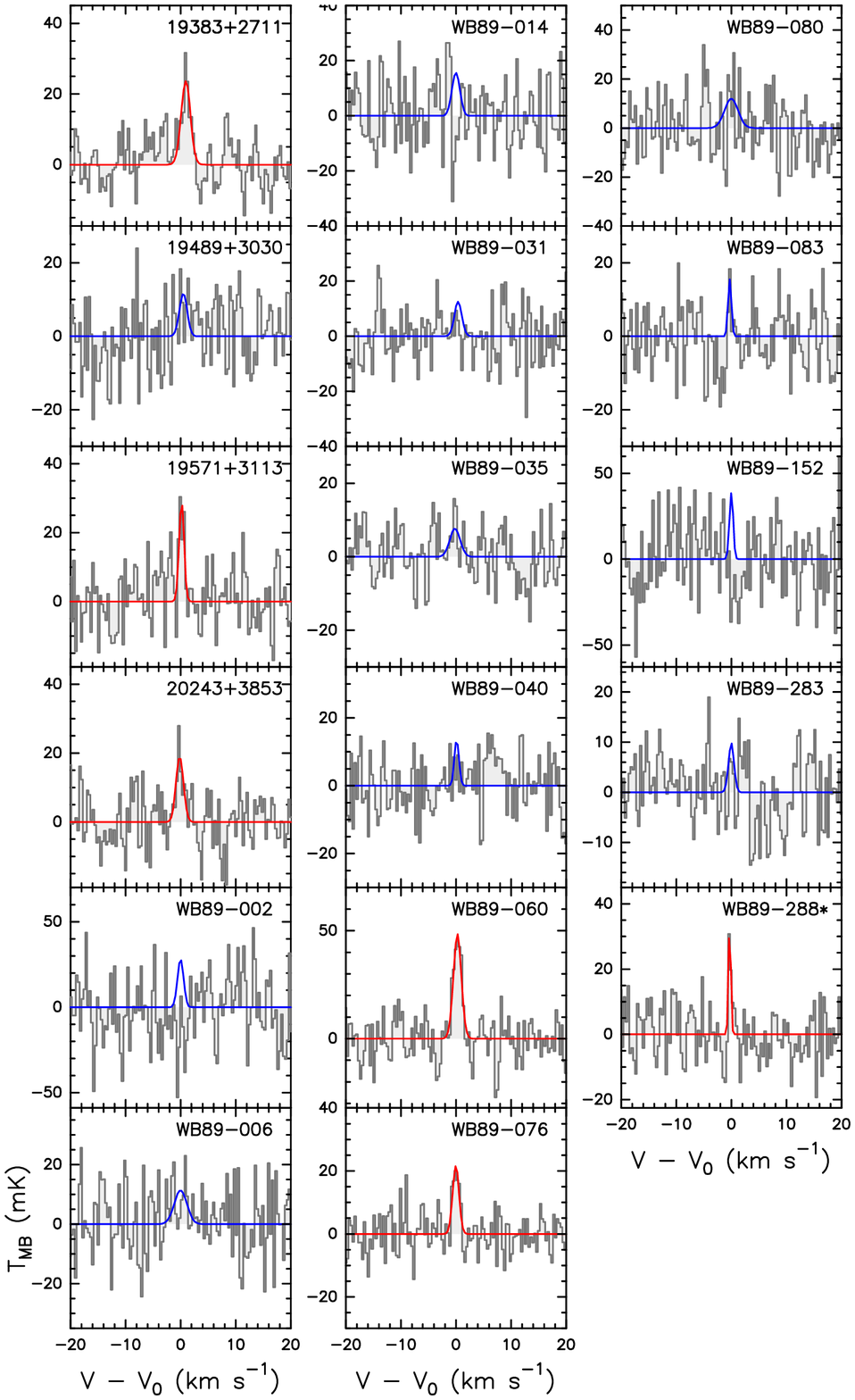}
\caption{Continued. }
\label{fig-spectra-hc15n-2}
\end{figure*}

\clearpage
\section{GCE models: Additional information on the \cratio\;ratio}

\label{sec:c-ratio-GCE}

\begin{figure}[htpb]
\centering
\includegraphics[width=22pc]{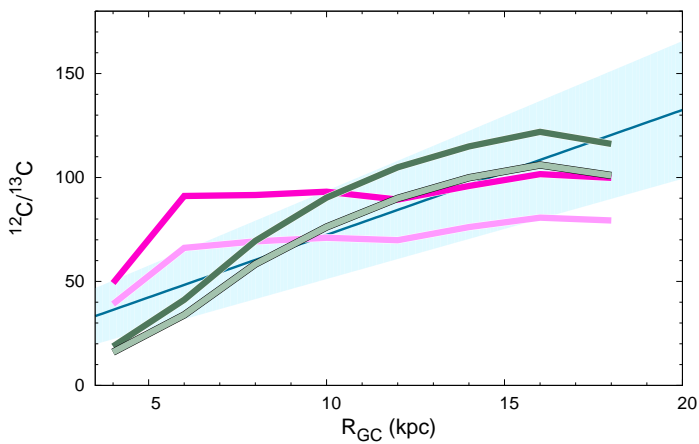}
\caption{Comparison between observed and predicted $^{12}$C/$^{13}$C ratios. The blue line indicates the observed $^{12}$C/$^{13}$C ratio trend along the Galactic disc by \cite{milam2005}, with its 1\,$\sigma$ uncertainty (light blue shaded area). The observed gradient is compared to the theoretical one obtained from different GCE models (see Table \ref{table-models}).}
\label{grad_mod3}
\end{figure}

In order to convert the $N$(H$^{13}$CN)/$N$(HC$^{15}$N) ratios to isotopic nitrogen ratios, we made use of the $^{12}$C/$^{13}$C gradient determined by \cite{milam2005}. 
As mentioned in Sect. \ref{res-results}, this trend was obtained for the inner Galaxy, and no observational constraints towards a sample of sources are available so far in the outer Galaxy.
In Fig.~\ref{grad_mod3} we show how the extrapolation of this trend up to $R_{\rm GC}$=20 kpc compares with the theoretical ratio predicted by the GCE models used in this work. When taking the respective uncertainties into account, a satisfactory agreement is found between observations and theoretical predictions.

We recall that in the GCE model adopted in this work $^{12}$C is produced as a primary element in both low- and intermediate-mass stars and massive stars, with the two sources contributing each about half of the solar $^{12}$C abundance (\citealt{romano2019}). The minor isotope, instead, has both a primary and a secondary origin: a large amount of primary $^{13}$C comes from massive fast rotators at low metallicities, while at higher metallicities secondary production in both low- and intermediate-mass stars and massive stars dominates. Our models also include a contribution to $^{13}$C synthesis from novae on long timescales. 

\section{Azimuthal variations}
\label{sec-azimutalvar} 

\begin{figure}[h!]
\centering
\includegraphics[width=20pc]{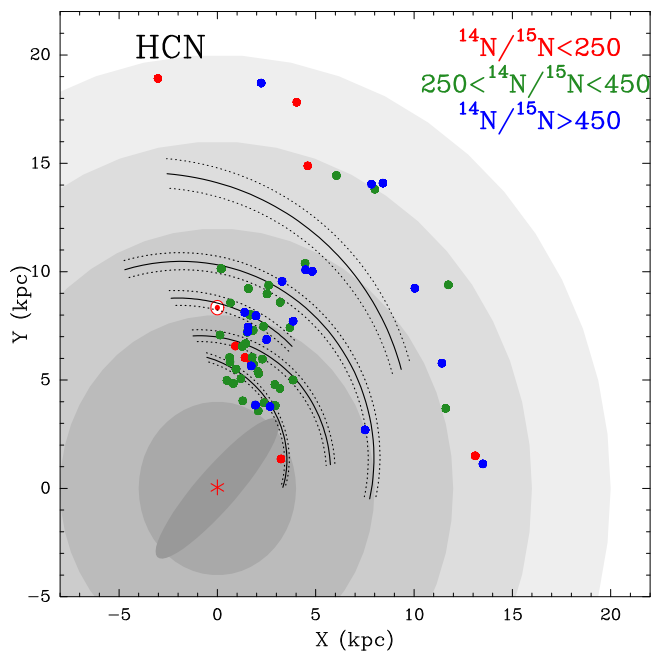}
\caption{Plan view of the Milky Way. The Galactic centre (red asterisk) is at (0,0), and the Sun (red Sun symbol) is at (0,8.34). The background grey discs
correspond to the Galactic bar region ($\sim$4 kpc), the solar circle ($\sim$8 kpc), co-rotation of spiral pattern ($\sim$12 kpc), and the edge of major star-formation regions ($\sim$16 kpc). The solid black lines indicate the centre of spiral arms traced by masers, and the dotted lines the 1$\sigma$ widths. For more details, see \citet{reid2014}. The filled circles represent the sources studied in this work and in \citet{colzi18b}, and the three colours are the \nratio\;ratios measured for HCN: in red values $<$ 250, in green values in between 250 and 450, and in blue ratios $>$ 450.}
\label{fig-azimuthal-vari}
\end{figure}
 
In this appendix we investigate a dependence with the Galactic longitude. Figure \ref{fig-azimuthal-vari} shows our results in the Galactic plane view of the Milky Way made by \citet{reid2014}. The different colours represent different \nratio\;ratios, as indicated by the figure labels. As already found in the inner Galaxy by \citet{colzi18b}, the new sources in the outer Galaxy do not present a clear trend along spiral arms or with the azimuthal angle.

\end{appendix}

\end{document}